\documentclass[useAMS,usegraphicx, usenatbib,fleqn]{mn2e}
\usepackage{natbib}
\usepackage{natbib}
\usepackage{amsmath,amssymb}
\usepackage{color}
\usepackage[dvips]{graphicx}
\usepackage{lscape}
\usepackage{booktabs}
\usepackage{longtable}
\usepackage{multicol}
\usepackage{multirow}

\usepackage[spanish]{babel}
\spanishdecimal{.}
\usepackage{times}
\DeclareMathOperator{\Likelihood}{\mathcal{L}}

\title[Environmental effects on the stellar mass-size relation]
{The effect of the environment on the stellar mass - size relation of present-day galaxies}
\author[Mar\'ia Cebri\'an and Ignacio Trujillo]{Mar\'ia Cebri\'an$^{1,2}$\thanks{E-mail:
mcebrian@iac.es} and Ignacio Trujillo$^{1,2}$\\
$^{1}$Instituto de Astrof\'isica de Canarias, V\'ia L\'actea s/n, E-38205 La Laguna, Tenerife, Spain\\
$^{2}$Departamento de Astrof\'isica, Universidad de La Laguna, E-38206 La Laguna, Tenerife, Spain}

\begin{document}

\renewcommand{\refname}{References}

\date{Accepted --. Received --; in original form --}

\pagerange{\pageref{firstpage}--\pageref{lastpage}} \pubyear{2013}

\maketitle

\label{firstpage}

\begin{abstract}

 To study how the environment can influence the relation between stellar mass and effective radius of nearby galaxies ($z<0.12$), we use a mass-complete sample extracted from the NYU-Value Added Catalogue. This sample contains almost $232000$ objects with masses up to $3\times10^{11}M_{\sun}$. For every galaxy in our sample, we explore the surrounding density within 2 Mpc using two different estimators of the environment. We find that galaxies are slightly larger in the field than in high-density regions. This effect is more pronounced for late-type morphologies ($\sim7.5\%$ larger) and especially at low masses ($M_*<2\times10^{10}M_{\sun}$), although it is also measurable in early-type galaxies ($\sim3.5\%$ larger). The environment also leaves a subtle imprint in the scatter of the stellar mass-size relation. This scatter is larger in low-density regions than in high-density regions for both morphologies, on average $\sim3.5\%$ larger for early-type and $\sim0.8\%$ for late-type galaxies. Late-type galaxies with low masses ($M_*<2\times10^{10}M_{\sun}$) show the largest differences in the scatter among environments. The scatter is $\sim20\%$ larger in the field than in clusters for these low-mass objects. Our analysis suggest that galaxies in clusters form earlier than those in the field. In addition, cluster galaxies seem to be originated from a more homogeneous family of progenitors.

\end{abstract}


\section{Introduction}
\label{sec:introduction}

The present-day stellar mass-size relation of galaxies hold information of the assembly history of galaxies across the cosmic time. Both the average size of the galaxies at a given stellar mass as well as the scatter of the relation are expected to reflect the evolutionary paths followed by the galaxies after their formation. These evolutionary tracks are considered to be different depending on the environment the galaxies inhabit. In fact, halos are expected to evolve fast and early in the highest density region of the Universe whereas in low-density environments this evolution is thought to be quieter \citep{sheth2004,gao2005,harker2006,maulbetsch2007}

Analysis of the luminosity-size and the stellar mass-size relations of high-z galaxies in the last 20 years \citep[e.g.][]{schade1996, lilly1998, simard1999, ravindranath2004, trujilloaguerri2004, mcintosh2005, barden2005, trujillo2004, trujillo2006, vanderwel2014} has shown that both late and early-type galaxies were more compact at a given mass or luminosity in the past. The size evolution is more dramatic for the early-type population than for the spiral-like galaxies \citep[see e.g.][]{trujillo2007, buitrago2008}. All these works, in combination with the decline in the number of compact galaxies in the nearby Universe \citep[see][]{trujillo2009,taylor2010, cassata2011, cassata2013, newman2012, szomoru2012, buitrago2013} prove that galaxies have undergone a significant size evolution with cosmic time.

Although the general evolution of the stellar mass-size relation with cosmic time (z $\lesssim$ 3) is starting to be well understood \citep[e.g.][]{stringer2013}, making sense of how the environment has affected the evolution of the stellar mass-size relation is by far less clear. For instance, at intermediate to high redshift ($0.5<z<2$), \cite{cooper2012, papovich2012} and \cite{delaye2013} find that elliptical galaxies have larger sizes when belonging to groups or clusters. However, works by \cite{rettura2010} and \cite{huertas2013a} showed no difference in a similar range of masses and redshifts. Moreover, even the opposite result has been found by \cite{raichoor2012}. Exploring early-type galaxies in clusters at z=1.3, they found that these objects are more compact in clusters than in the field. An interesting result by \cite{lani2013} reported that although early-type galaxies are larger when residing in dense environments at z $>$ 1, this trend seems to be weakened as we come closer to the present-day Universe (z $<$ 1).

In the nearby Universe, the effect of the environment in the stellar mass-size relation of the galaxies is also not clear. In the case of early-type galaxies, \cite{maltby2010}, \cite{huertas2013b} and \cite{cappellari2013} found no size difference between those objects with the same mass in different environments. However, \cite{poggianti2013} found that elliptical galaxies are more compact in clusters than in the field. For spiral galaxies, \cite{maltby2010} report a slight trend for low/intermediate mass ($10^9M_{\sun}<M_*<10^{10}M_{\sun}$) spirals to be larger in the field than in groups. \cite{fernandez2013} using the AMIGA sample detected that massive ($M_*>10^{10}M_{\sun}$) isolated spirals are $20\%$ larger than those located in dense environments, however, no size differences were reported for lower mass spirals.

Both in the nearby Universe as well as in the high-z galaxies, the discrepancies found in the literature can be explained by the relatively modest sample of galaxies analysed, typically on the order of few thousands of galaxies or less, with the largest sample being that used in the work of \cite{huertas2013b} which includes $\sim12000$ galaxies. One of the aims of this paper is to re-analyse this situation by exploring the stellar mass-size relation of both early and late-type galaxies using the large collection of data available in the NYU-VAGC Catalogue \citep{blantonNYU}. This huge dataset allows the sizes of the galaxies to be analysed at a fixed stellar mass with an unprecedented quality from the statistical point of view. In fact, the number of galaxies we use in this work is a factor $\sim50$ larger compared to \cite{maltby2010} and  \cite{fernandez2013}.

In addition to the average size of galaxies at a given stellar mass depending on the environment, another quantity that deserves attention is the scatter of this relation. It is worth noting that the measured scatter found in the Fundamental Plane and other scaling relations \citep{nipoti2009, nair2011,bezanson2013} is much lower than the one predicted theoretically \citep{nipoti2003,nipoti2009,nipoti2012,ciotti2007,shankar2010b,shankar2013}. The large scatter in the theoretical works are the consequence of the intrinsic stochastic nature of the merging processes. For this reason,  only models with very fine-tuned input conditions on the progenitors are able to reproduce the low dispersion values observed in the stellar mass-size distribution. Despite the enormous interest of measuring this quantity, little effort has been made observationally in this direction. To the best of our knowledge, using Sloan Digital Sky Survey (SDSS) data, only \cite{shen2003} have quantified this dispersion segregating the galaxies according to their morphology in the stellar mass-size plane. However, no attempt has been made measure the scatter segregating the galaxies according to their environment. This is understandable again due to the modest size of the samples used to explore the effect of the environment both at low and high-z. As we will show through this paper, the environment only marginally affects the scatter of the stellar mass-size relation and, consequently, the use of large datasets is necessary to identify the role played by the environment in shaping the stellar mass-size plane.

This paper is organized as follows: Section \ref{sec:data} contains a description of the data used in this work; Section \ref{sec:environment} describes the different definitions of environment and the sample selection; Section \ref{sec:size-scatter} is dedicated to study the stellar mass-size plane. Finally, our results are shown in Section \ref{sec:results} and discussed in Section \ref{sec:discussion}. A summary can be found in Section \ref{sec:summary}. Throughout the paper a standard $\Lambda CDM$ cosmology is adopted: $\Omega_M=0.3$, $\Omega_{\Lambda}=0.7$ and $H_0=70$ km s$^{-1}$Mpc$^{-1}$.

\section{Data and sample selection}
\label{sec:data}

Our data is entirely drawn from the publicly available NYU Value-Added Galaxy Catalogue \citep{blantonNYU} based on the SDSS-DR7 \citep{sdss-dr7}. This catalogue contains around $2.5\times10^6$ objects with spectroscopic redshift determination, mostly below $z\sim0.25$. Besides photometric and spectroscopic information extracted from SDSS-DR7, the NYU-VAGC catalogue also provides structural parameters, K-corrections and stellar mass estimates \citep{blantonKcorrect} derived using a \cite{chabrier2003} Initial Mass Function (IMF) with population synthesis model from \cite{Bruzual&Charlot}.

In this work, to minimize potential biases due to the size evolution of the galaxies with redshift (e.g. \citealt{trujillo2007,buitrago2008}) we restrict our redshift range to $z=0.005-0.12$, which is equivalent to a cosmic time interval of only 1.5 Gyr. We have explored whether there is any significant size evolution within this redshift range. To do that, we have divided our sample in two redshift bins ($0.005<z<0.076$ and $0.076<z<0.12$) and computed the mean sizes within each of these bins. There is not size difference between these two intervals.

\begin{figure}
  \includegraphics[width=84mm]{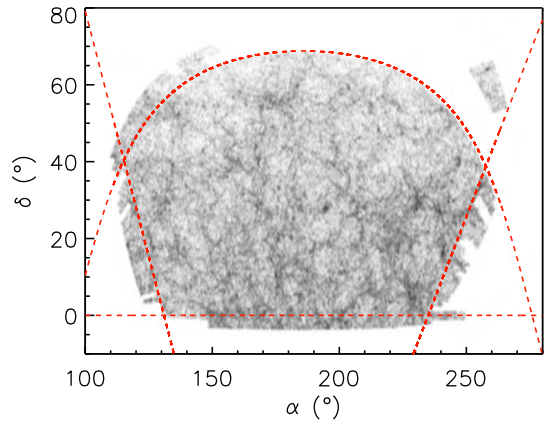}
  \caption{Projected distribution of the galaxies in the NYU-VAGC catalogue used in this work. The red-dashed lines indicate the limits of the area surveyed to estimate the environmental densities.}
  \label{fig:geometria}
\end{figure}

The SDSS-DR7 catalogue has irregular limits (see Figure \ref{fig:geometria}). At the edges of the survey there is a potential bias affecting the definition of environment when the surrounding objects are used as an indicator of the galaxy density. This could create artificially low-density regions. To simplify the treatment of this problem, a volume trimming following \cite{varela2012} is carried out by limiting the projected area as follows.

\begin{itemize} \label{eq:geometry}
 \item Southern limit: $\delta>0^o$
 \item Western limit: $\delta>-2.555556(\alpha-131^o)$
 \item Eastern limit: $\delta>1.70909(\alpha-235^o)$
 \item Northern limit: $\delta<\arcsin[\frac{0.93232\sin(\alpha-95.9^o)}{\sqrt{1-[0.93232\cos(\alpha-95.9^o)]^2}}]$
\end{itemize}

\subsection{Stellar mass completeness of the sample}
\label{sec:completeness}

In order to properly analyse the stellar mass-size relation in different environments, we need to select a mass-complete sample. The NYU-VAGC catalogue is spectroscopically complete ($\sim99\%$) up to $r\sim17.7$ magnitude \citep{strauss2002}. This flux cut in the \textit{r}-band introduces a bias in the stellar mass distribution of the galaxies that needs to be addressed before exploring the stellar mass-size distribution. In particular, this effect must be corrected to avoid our sample being biased towards young galaxies, especially at lower masses.

\begin{figure}
  \includegraphics[width=84mm]{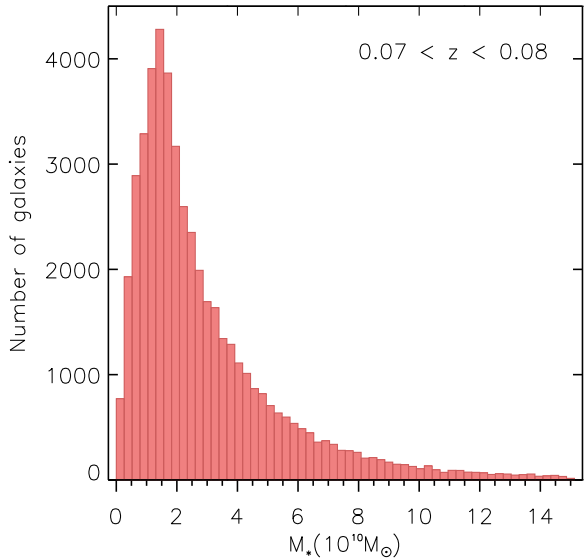}
  \caption{Stellar mass histogram for galaxies in our sample with redshift $0.07<z<0.08$. The number of detected objects increases as we move towards lower stellar masses until the maximum of the histogram is reached. At this point, although the number of galaxies should keep growing, less massive objects start to be missed because of the \textit{r}-band spectroscopic magnitude limit ($r\sim17.7$) from SDSS-DR7.}
  \label{fig:completeness_histo}
\end{figure}

\begin{figure}
  \includegraphics[width=84mm]{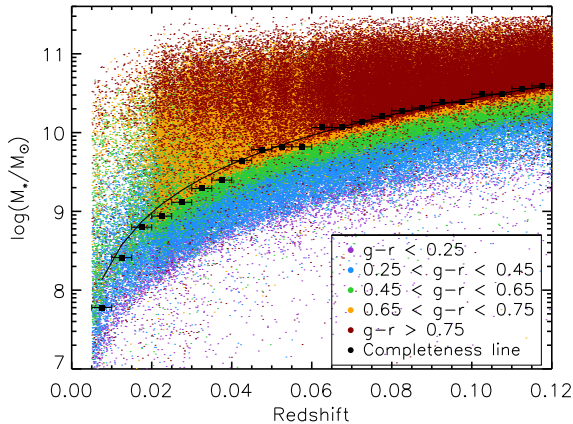}
  \caption{Stellar mass and redshift distribution of the NYU-VAGC catalogue used in this work. Different dot colours represent different g-r colours for the galaxies as stated in the legend. Black solid squares indicate the completeness mass limit based on the peak of the stellar mass histogram at each redshift (see Figure \ref{fig:completeness_histo}). The black line is the fit to the completeness mass limit as explained in the text.}
  \label{fig:completeness}
\end{figure}

To illustrate how the \textit{r}-band magnitude limit affects the stellar mass distribution of our galaxies, we show in Figure \ref{fig:completeness_histo} the distribution of mass in the redshift range $0.07<z<0.08$. Starting from the higher mass bins, the number of galaxies increases as we move towards the lower masses until a peak in the histogram is reached. At this point, galaxies with masses below that peak start to be undetected and the sample becomes incomplete. Consequently, we identify the mass where the peak of this histogram is reached as the mass completeness limit. As it depends on the redshift range, we divide our sample in several redshift bins and study the corresponding histogram for each of them, obtaining the completeness limits showed in Figure \ref{fig:completeness}. The black squares are the stellar mass above which the sample is mass complete and the horizontal error bars represents the redshift bin width. At the high redshift limit of our sample ($z=0.12$), the corresponding mass 
limit is $4\times10^{10}M_{\sun}$.

Once we have determined the stellar mass limit for different redshift bins, we can fit the mass completeness limit using an analytical expression with the form:

\begin{equation} \label{eq:completness}
  \log \left[M_*(z)/M_{\sun}\right]=0.4\left(-M_r(z)+2.5\log(M/L)_{r,limit}+M_{sun,r}\right)
\end{equation}

where $M_*(z)$ is the stellar mass limit as a function of the redshift in solar masses, $(M/L)_{r,limit}$ is a free parameter of the fit, corresponding to the mass-to-light ratio in the \textit{r}-band at the completeness mass limit, $M_{sun,r}$ is the absolute magnitude of the sun also in the \textit{r}-band equal to $4.68$ mag and $M_r(z)=r-5\log(D_L(z))+5$ is the absolute magnitude limit of the survey as a function of the redshift, corresponding to $r=17.7$ magnitudes. $D_L(z)$ represents the luminosity distance in pc for an object at redshift $z$. The fit is shown in Figure \ref{fig:completeness}. The fitting  of our experimental mass limits gives us a value for the mass-to-light in the \textit{r}-band ratio of $(M/L)_{r,limit}=2.059$. This value of the mass-to-light ratio corresponds to a population of $5.6$ Gyr, according to the MIUSCAT SEDs developed by  \cite{vazdekis2012} and \cite{ricciardelli2012} with a \cite{kroupa2001} IMF and solar metallicity. The results of this analysis show that, even selecting the peak of the mass distribution at each redshift, it is likely that some very old and metal-rich galaxies could be missed from our sample. In order to test whether our results could be affected by this potential incompleteness, we have conducted the full analysis of our work using a more conservative mass-to-light ratio of $(M/L)_{r,limit}=4$, i.e. an age of 12.5 Gyr. That corresponds to stellar mass limit of $7.5\times10^{10}M_{\sun}$ at $z=0.12$. Our tests show that using this very conservative mass limit do not change our results while our sample of galaxies would be considerably reduced and henceforth our errors increased. Consequently, in what follows we use the stellar mass limit determined using Equation \ref{eq:completness} and equivalent to $(M/L)_{r,limit}=2.059$.

The effect of the age on the completeness of the stellar mass is well illustrated in Figure \ref{fig:completeness}. Young (blue) galaxies are brighter than old (red) galaxies of the same stellar mass. Consequently, they are overrepresented when the samples are drawn from an apparent magnitude limit (in this case in the \textit{r}-band). This shows the importance of selecting a truly mass complete sample. To illustrate this point and assure that our criteria for completeness also account for the colour, Figure \ref{fig:completeness} shows galaxies in the stellar mass-redshift plane, segregating each galaxy according to its \textit{g-r} colour. This segregation shows that, at a given redshift, the galaxies with the lower stellar masses are only blue, whereas for stellar masses above the completeness limit we have the two populations of galaxies.

After this consideration, our mass-complete sample contains $231905$ objects. Our final sample is shown in Figure \ref{fig:neigh-completeness}.

\subsection{Size and morphology determination}
\label{sec:size}

To construct the stellar mass-size distribution of our sample, a reliable measure of the size of each galaxy is crucial. The sizes of the galaxies as well as the shape of their surface brightness distributions are provided by the NYU-VAGC catalogue. These parameters were determined using a \citet{sersic1968} fit to the radial intensity profile obtained using circular apertures \citep{blanton2005}:

\begin{equation} \label{eq:sersic}
  I(r)=I(0)\rmn{exp}[-b_n(\rmn{R}/\rmn{R_e})^{1/n}]
\end{equation}

$I(0)$ is the central intensity, $n$ is the S\'ersic index and $b_n$ is a function of $n$ such as $\rmn{R_e}$ is the effective radius. The S\'ersic index correlates with the morphology of the galaxy \citep{andredakis1995}: objects with $n<2.5$ are mostly discy, while those with $n>2.5$ are bulge-dominated or spheroids \citep{ravindranath2004}. We take advantage of this phenomenon to separate our sample into galaxies with different morphologies ($n>2.5$ and $n<2.5$) and probe the effect of the environment in each subgroup. In this work, we use the circularized effective radius $R_e$ as a measurement of the size of the galaxy.

\section{Characterizing the environment of the galaxies}
\label{sec:environment}

In this paper we study the stellar mass-size relation in different large scale ($\sim2$ Mpc) environments. We use two different approaches to define whether the galaxies reside in a high-density or a low-density region. Our first estimation of the environment is computed using the stellar mass surrounding each object within a physical radius. For our second estimator of the environment, we use several catalogues of galaxy clusters. Galaxies in these clusters are compared with galaxies in the field. In this section, we detail how these different density indicators are obtained.

\subsection{First environment characterization: galaxy number density}
\label{sec:ndensity}

Different methods for measuring the environment of galaxies are used in the literature. These methods can be divided in two main groups: estimators based on the number of neighbours within an aperture of a fixed radius and estimators based on the distance to the $n$-th nearest neighbour. \cite{cooper2005} investigated the effects of survey edges, redshift-space distortions (the 'finger of God' effect) and other effects on mock catalogues among different environmental estimators, finding that fixed aperture methods are more robust to the previously mentioned effects while providing a direct density measure. Despite its advantages, this method present two drawbacks that have to be taken into account: the sensitivity in low-density environments and the fact that is a quantized measurement. Other studies as those of \cite{haas2012} and \cite{muldrew2012} also support fixed aperture methods as a useful tool to define the surrounding density of a galaxy. In particular, \cite{haas2012} found that the optimal 
radii for this kind of method is $\sim2$ Mpc.

\begin{figure}
  \includegraphics[width=84mm]{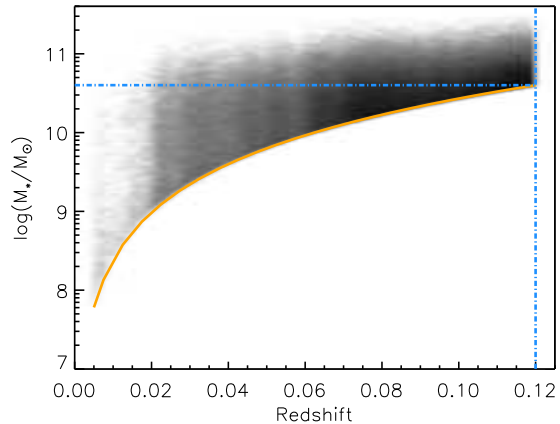}
  \caption{The redshift-mass plane of the stellar mass-complete galaxies in the NYU-VAGC catalogue. The solid orange line shows the mass completeness line of the sample. The vertical and horizontal blue lines indicate the redshift and the mass limit used to explore the density of different environments. In order to lighten the plot, the density of objects is represented as a shaded surface instead of using individual points.}
  \label{fig:neigh-completeness}
\end{figure}

With these studies in mind, we use a fixed aperture method to make our first measurement of the environmental density. The characterization of the environmental density is done as follows. For each galaxy in our stellar mass-complete sample, we explore what is the number and total mass of the surrounding galaxies in a sphere of 2 Mpc radius, centred on our targeted galaxy. Note, however, that not all the surrounding galaxies can be used to make an estimation of the environmental density around our galaxies. In fact, to conduct an homogeneous description of the environment along our whole sample, only galaxies with $M_*>4\times10^{10}M_{\sun}$ are considered to estimate the density (see Figure \ref{fig:neigh-completeness}). This mass limit corresponds to the mass limit of our sample at $z=0.12$.

To define the sphere and the position of each object in the survey, we use cartesian coordinates defined by the set of Equations \ref{eq:cartesian}, where $\alpha$ is the Right Ascension of the galaxy, $\delta$ its Declination, $z$ its redshift and $D(z)$ is the co-moving radial distance:

\begin{equation} \label{eq:cartesian}
\begin{array}{l}
  X=D(z)\cos\delta\cos\alpha \\
  Y=D(z)\cos\delta\sin\alpha \\
  Z=D(z)\sin\delta \\
\end{array}
\end{equation}

Our density $\rho$ for each galaxy is hence defined as:

\begin{equation} \label{eq:density}
 \rho_i=\frac{1}{\frac{4}{3}\pi R^3}\sum_{k}^{N}{M_{i,k}}
\end{equation}

where $M_{i,k}$ is the stellar mass of the $k$-th neighbour with $M_*>4\times10^{10}M_{\sun}$ located at a radial distance less than 2 Mpc to the $i$-the galaxy in our sample. In the above equation $R=2$ Mpc.

\begin{figure}
  \includegraphics[width=84mm]{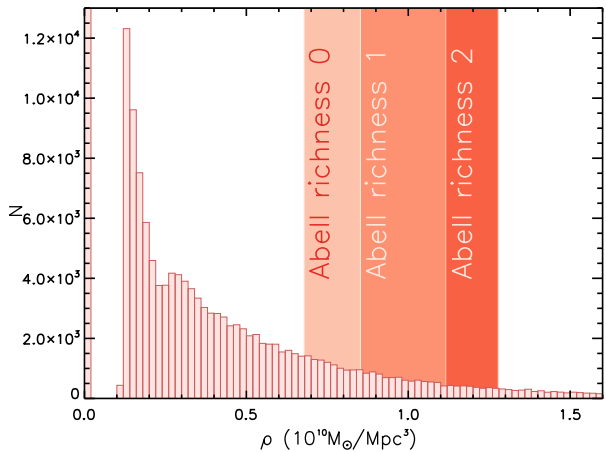}
  \caption{Distribution of the environmental density of our galaxies using our criteria defined in section \ref{sec:ndensity}. To make sense of our numbers the coloured regions indicates the typical density values of galaxies located in Abell clusters of different richness.}
  \label{fig:histogram}
\end{figure}

Figure \ref{fig:histogram} represents the distribution of the environmental density for the galaxies in our sample. The gap between the bin at $\rho=0$ and the first at $\rho\sim0.12$ reflects the discretization of our estimator, as previously mentioned: the minimum non-zero value $\rho$ can take is that corresponding to one companion with a mass of $4\times10^{10}M_{\sun}$. To check if our density estimator truly segregates the different large-scale structures, we computed the same number $\rho$ for galaxies in our catalogue located in Abell clusters \citep{abell1989}. When we compare their values with the histogram in Figure \ref{fig:histogram}, it is clear that objects belonging to clusters are in the highest-density regions of the histogram.

To assess the effect of the environment on the stellar mass-size distribution we have considered the galaxies in the two most extreme ends of Figure \ref{fig:histogram}. Galaxies in the top $10\%$ (i.e. $\rho>0.76$) of the distribution are considered to be located in high-density regions, whereas galaxies in the bottom $10\%$ reside in low-density regions. Since more than $10\%$ of the main sample have $\rho=0$, the bottom $10\%$ have been randomly taken from all the galaxies in our sample with $\rho=0$ (first density bin of the distribution on Figure \ref{fig:histogram}).

Several tests have been run changing the aperture $R$ used in the search and the percentage (i.e. the value of $\rho$) defining each environment. As we will show later, these tests show that these changes do not severely affect the final result. The present definition provides a compromise between exploring extreme environments and getting a statistically significant representation for each of them.

\subsection{Second environment characterization: Cluster galaxies vs. field galaxies}
\label{sec:clusters}

Galaxy clusters are the highest-density large-scale structures in the Universe. Consequently, another way to compare galaxies in different environments is by selecting galaxies residing in galaxy clusters and confronting them with field galaxies (those outside these cluster environments). In order to do that, we compile a large sample of galaxy clusters in our explored volume using different catalogues: the Abell catalogue \citep{abell1989}, a catalogue extracted from SDSS-DR6 \citep{szabo2011}, three catalogues from SDSS III \citep{einasto2012,tempel2012,wen2012}, the GMBCG cluster catalogue \citep{hao2010} and the XMMi-SDSS galaxy cluster survey \citep{takey2011}. A total of 1877 galaxy clusters are used in this work, containing more than $12600$ galaxies within our stellar mass-complete subsample.

Similarly to the previous estimator, a sphere of 2 Mpc radius is defined around the centre of each cluster using cartesian coordinates as in section \ref{sec:ndensity}. Thus, every galaxy inside this sphere is considered a cluster galaxy and the rest of the sample form the field sample. To avoid duplicates due to overlapping, if a galaxy is selected as belonging to two different clusters, the object is only assigned to the nearest cluster.

\section{Size and scatter of the stellar mass-size distribution}
\label{sec:size-scatter}

The size distribution of the galaxies at a fixed stellar mass follows a log-normal distribution (e.g. \citealt{shen2003}) given by:

\begin{equation} \label{eq:log-normal} 
 f(R,\overline{R}(M),\sigma_{\ln R}(M))=\frac{1}{\sqrt{2\pi}\sigma_{\ln R}(M)R}exp\left\{-\frac{\ln^2[R/\overline{R}(M)]}{2\sigma_{\ln R}^2(M)}\right\}
\end{equation}

where $\overline{R}(M)$ is the mean of the distribution and provides us with the typical size for a galaxy given its mass, while $\sigma_{\ln R}(M)$ is the dispersion of the distribution and indicates how concentrated are the objects around the typical radius value. In our work, we assume that the same expression used in Equation \ref{eq:log-normal} is valid for describing the size distribution of galaxies in different environments. Consequently, we will explore whether the parameters that characterize the size distributions are different depending on where the galaxies reside. 

\subsection{Parameters estimation: Maximum likelihood method}
\label{sec:par-estimation}

In order to accurately estimate the parameters $\overline{R}(M)$ and $\sigma_{\ln R}(M)$, we use a maximum likelihood method. To implement this method, we consider a sample of $n$ galaxies within a mass range and compute the likelihood for each possible value of $\overline{R}(M)$ and $\sigma_{\ln R}(M)$ as follows:

\begin{equation} \label{eq:likelihood_prod} 
 \Likelihood(\overline{R}(M),\sigma_{\ln R})=\prod_{i=0}^n\frac{1}{\sqrt{2\pi}\sigma_{\ln R}R_i}exp\left\{-\frac{\ln^2[R_i/\overline{R}(M)]}{2\sigma_{\ln R}^2}\right\}
\end{equation}

 where $R_i$ represents the size of each of the $n$ galaxies in the sample. The above expression is more easily computed if logarithms are taken on both sides, turning the product into a sum:

\begin{equation} \label{eq:likelihood_sum} 
 \ln\left[\Likelihood(\overline{R}(M),\sigma_{\ln R})\right]=\sum_{i=0}^n\left(\ln\left[\frac{1}{\sqrt{2\pi}\sigma_{\ln R}R_i}\right]-\frac{\ln^2[R_i/\overline{R}(M)]}{2\sigma_{\ln R}^2}\right)
\end{equation}

The best estimates for the parameters ($\overline{R}(M)$, $\sigma_{\ln R}(M)$) will be those producing the maximum value of $\Likelihood$. The maximum likelihood estimation is implemented here using an iterative grid search in the parameter space, obtaining the most likely values for our parameters given our distribution. With this method, we also obtain their 1$\sigma$-errors, computed as the contour including $68.2\%$ of the normalized likelihood in the grid.

\subsection{Mean size and dispersion as a function of the environmental density $\rho$}

\begin{figure*}
  \includegraphics[width=168mm]{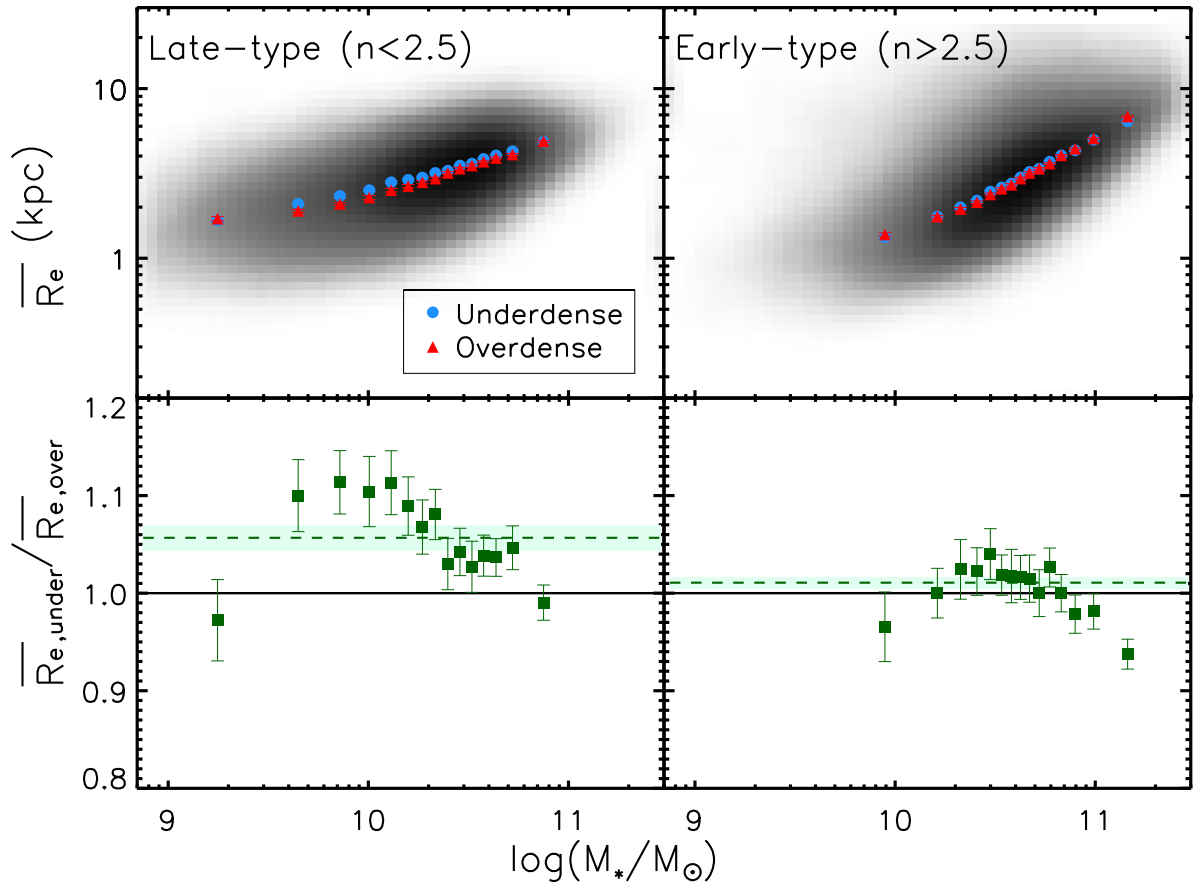}
 \caption[caption]{The stellar mass-size relation and their differences for different environments and morphologies. Upper panels show the overall distributions for discy and spheroid-like objects as a shaded surface. Over-plotted in those distributions are the mean size of the galaxies in the $10\%$ lowest density (blue filled circles) and the $10\%$ highest density regions (red filled triangles). Lower panels show the ratio between the mean sizes in the most underdense and overdense samples (error bars representing 1$\sigma$-errors). The green dashed line is a fit to all the distribution of points and indicates the robust mean\footnotemark[1] with 1-$\sigma$ error represented as green shadowed area.}
  \label{fig:diffR-neigh}
\end{figure*}

\begin{figure*}
  \includegraphics[width=168mm]{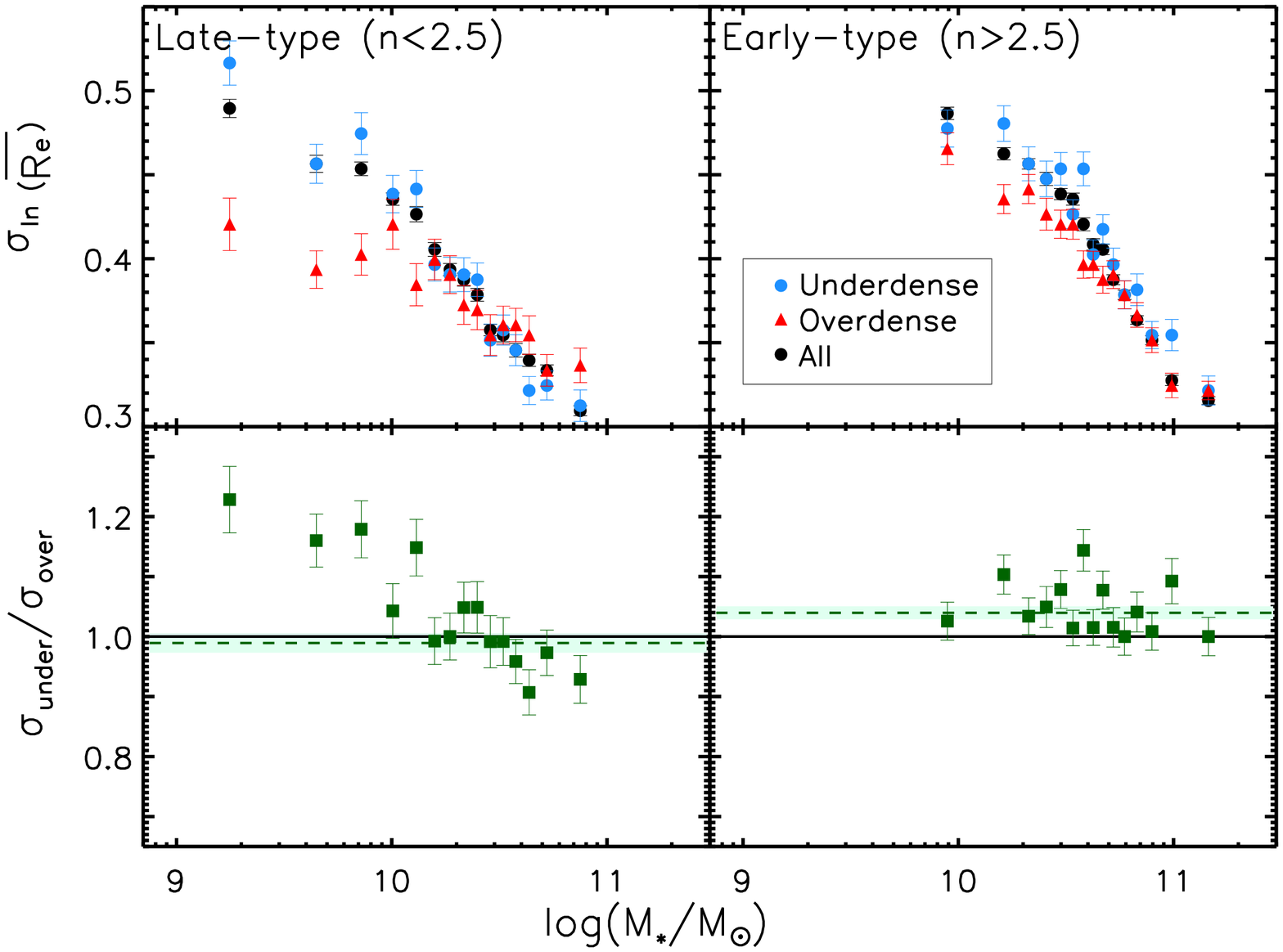}
 \caption[caption]{Upper panels: Scatter of the size distribution for different mass bins. Galaxies are segregated according to the density where they inhabit (blue filled circles and red filled triangles). Lower panels: Ratio between scatters of low-density and high-density region as well as their 1$\sigma$-errors (green squares). The green dashed line show the robust mean of all the points in the distribution and its 1-$\sigma$ error is plotted as a green filled area.}
  \label{fig:diffS-neigh}
\end{figure*}

To study how the galaxies are distributed in the stellar mass-size plane, we segregate them by their morphology (estimated using the S\'ersic index as pointed in section \ref{sec:size}): objects with $n>2.5$ are considered spheroidal while objects with $n<2.5$ are mostly disc-like galaxies.

Having the galaxies separated according to their morphology, we apply the maximum likelihood estimation to obtain the most likely values for the mean size and the scatter of the distribution. In order to do that, we divide each morphological and environmental sample in different mass bins. The width of each mass bin is taken so that each mass bin contains the same number of galaxies from the main sample. For each bin of stellar mass we use the maximum likelihood method to estimate both the more likely mean size $\overline{R}(M)$, dispersion $\sigma_{\ln R}(M)$ and their 1$\sigma$-errors. 

Figure \ref{fig:diffR-neigh} represents the mean size estimation for the different morphologies and environments. In the lower panels of Figure \ref{fig:diffR-neigh}, we show the ratio between $\overline{R}(M)$ for the $10\%$ most underdense and $10\%$ most overdense regions, both for late ($n<2.5$) and early-type ($n>2.5$) galaxies. Late-type galaxies (left panels in Figure \ref{fig:diffR-neigh}) with intermediate masses $M_*<2\times10^{10}M_{\sun}$ show the largest difference among environments, being larger when they reside in low-density regions. On average, late-type galaxies in low-density environments are $6\pm1\%$ larger than those in the highest density regions. Regarding early-type galaxies (right panels in Figure \ref{fig:diffR-neigh}), they present also a difference of size with the environment, although this is less prominent than for the late-type case: $1.1\pm0.6\%$.

In addition to the mean sizes, we analyse the dispersion of the distribution in different environments. Our results are shown in Figure \ref{fig:diffS-neigh}. The most remarkable finding is that for early-type galaxies the scatter in the distribution of sizes is $4\pm1\%$ larger in the underdense region compared to overdense zones. For late-type galaxies there is no mean difference among the environments (i.e. a mean difference of $1\pm2\%$ when the full population is considered). However, it seems that there is a trend towards larger scatter of the sizes in the low mass ($M_*<10^{10}M_{\sun}$) regime of galaxies in underdense regions and the opposite behaviour at larger masses ($M_*>10^{10}M_{\sun}$). These differences could be as high as $20\%$, depending on the exact bin mass.

We summarize our results of this section in Tables \ref{table:lateneigh} and \ref{table:earlyneigh}. As previously explained, the 1-$\sigma$ errors were obtained as the contour in the parameter space which contains $68.2\%$ of the normalized likelihood.

\footnotetext[1]{The robust mean provides a measure of the mean trimming away outliers in the distributions. In this work, the \textsc{IDL} routine \textsc{resistant\_mean.pro} is used for this purpose, trimming points which lie outside the 2$\sigma$ confidence interval.}

\subsection{Mean size and dispersion for field and cluster galaxies}

\begin{figure*}
  \includegraphics[width=168mm]{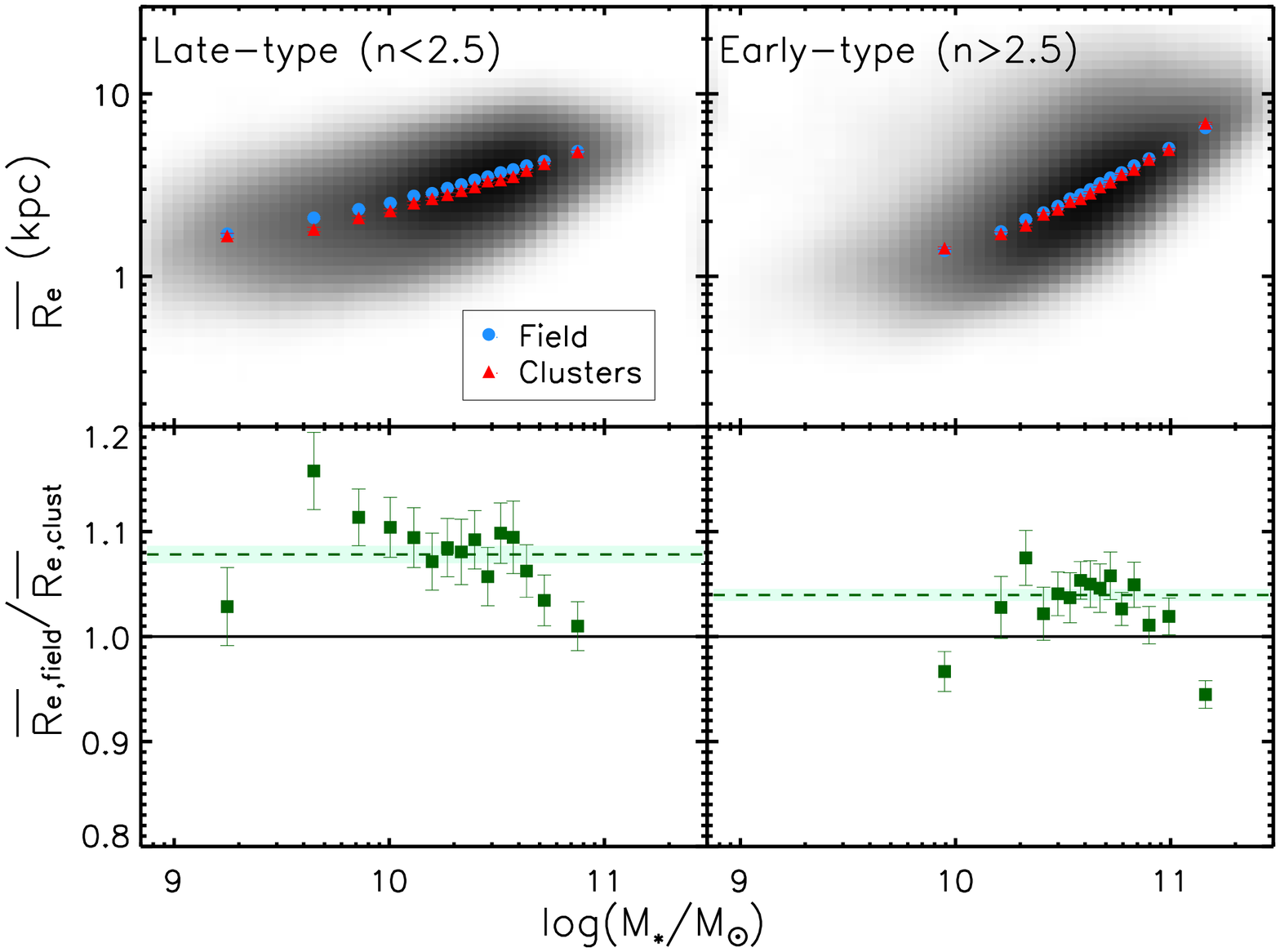}
 \caption{Same as Figure \ref{fig:diffR-neigh} but using samples corresponding to galaxies in clusters and in the field. The upper panels show the overall distributions for discy and spheroid-like objects (shaded surface). Overplotted on those distributions are the mean size of the galaxies in the field (blue filled circles) and in the cluster regions (red filled triangles). The lower panels show the ratio between sizes in the clusters and in the field as green filled squares. Errors bars represent 1$\sigma$-errors and the mean of these ratios is the green dashed line with its $1\sigma$-error indicated with a green filled area.}
  \label{fig:diffR-clusters}
\end{figure*}

\begin{figure*}
  \includegraphics[width=168mm]{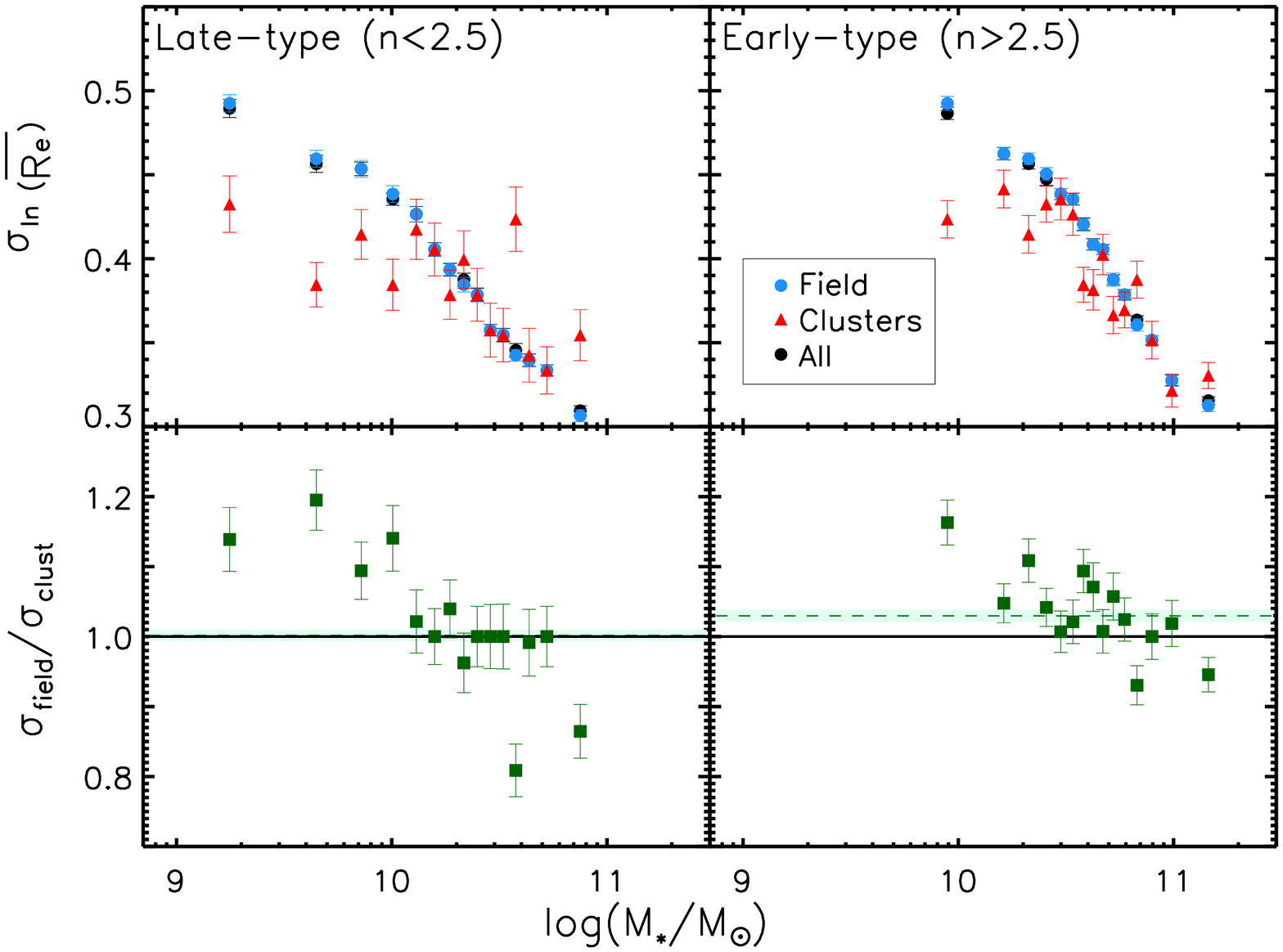}
 \caption{Same as Figure \ref{fig:diffS-neigh} but using samples for galaxies in clusters and in the field. The upper panels represent the scatter of the size distribution for different mass bins. Galaxies are segregated according to the density where they reside (blue filled circles and red filled triangles). The lower panels show the ratio among the scatter of the galaxies in the field and in the clusters (green filled squares) as well as their 1$\sigma$-errors. The robust mean of these differences is indicated with a green dashed line and a green filled area for its $1\sigma$-error.}
  \label{fig:diffS-clusters}
\end{figure*}

We have repeated the same analysis as before but for cluster and field galaxies. Cluster and field galaxies are segregated according to their morphology given by the S\'ersic index (as explained in section \ref{sec:size}). For each morphology, the main sample is further divided in different mass bins, each of them containing the same number of galaxies, as in the previous subsection. We compute the best estimations for $\overline{R}(M)$ and $\sigma_{\ln(R)}(M)$  using the maximum likelihood method explained in section \ref{sec:par-estimation} for each mass bin.

The stellar mass-size relation for galaxies in the field and in clusters can be seen in Figure \ref{fig:diffR-clusters}. This figure shows how the effects of the environment are different depending on the morphology: late-type galaxies are on average $7.8\pm0.6\%$ larger in the field than in the clusters, while early-type galaxies present a similar trend but with smaller significance, being objects inhabiting clusters slightly larger ($4.0\pm0.8\%$) than their counterparts in the field.

The dispersion of the stellar mass-size distribution is shown in Figure \ref{fig:diffS-clusters}. For late-type galaxies the mean of the ratio of scatters among the environments is small: $0.2\pm0.8\%$.  Despite this, and in agreement with our first environment characterization, low mass objects ($M_*<10^{10}M_{\sun}$) residing in the field show a stellar mass-size distribution with larger scatter than their counterparts residing in clusters, but there seems to be no trend for the high mass end ($M_*>10^{10}M_{\sun}$).

Early-type galaxies show a trend of being more scattered on the field that in clusters, with an average of $3.0\pm0.9\%$. Again, this trend is more pronounced for the lowest mass bins ($M_*<2\times10^{10}M_{\sun}$).

Numerical values for the parameters obtained and their errors are shown in tables \ref{table:lateclust} and \ref{table:earlyclust} with the corresponding 1$\sigma$-errors.


\section{Results}
\label{sec:results}

The significant increase in the number of galaxies provided by the SDSS survey allows us to explore the stellar mass-size distribution of galaxies as a function of the environment with an unprecedented accuracy. The mean differences in the effect of the environment found in our work are summarized in Table \ref{table:mean_results}, while the detailed results for each stellar mass are given in Tables \ref{table:lateneigh} to \ref{table:earlyclust}. 

\begin{table*}
    \begin{tabular}{ccccccc}
      \hline
	  & $\overline{R}_{under 10\%}/\overline{R}_{over 10\%}$  &$\overline{R}_{field}/\overline{R}_{clust}$ & $\sigma_{under 10\%}/\sigma_{over 10\%}$&  $\sigma_{field}/\sigma_{clust}$\\
      \hline
	\textbf{Late-type} &	$1.057\pm0.013$	&   $1.078\pm0.006$	& $0.989\pm0.016$	&	$1.002\pm0.008$\\
	\textbf{Early-type} &	$1.011\pm0.006$	&   $1.040\pm0.008$	& $1.04\pm0.01$		&   	$1.030\pm0.009$\\
        \end{tabular}
   \caption{Differences in the mean size and scatter of the galaxies depending on the environment inhabited and their morphology.}
     \label{table:mean_results}
\end{table*}

Although using the top and lower $10\%$ of our density distribution allows us to obtain high-statistics samples, it is worth studying the impact of considering a more extreme definition of environment in our results. Table \ref{table:mean_results5} shows the same results as Table \ref{table:mean_results} but using the densest $5\%$ (i.e. $\rho>1.05$) and the less dense $5\%$ of our density distribution as representative for the low and the high-density environments respectively. On average, taking more extreme densities increases the differences both in mean sizes and scatter among the different environments, showing an effect very similar to that obtained using galaxy clusters. This leads us to think that the higher the value of the density used, the better is the correlation with the cluster environment. Using more relaxed conditions for defining the different environments (e.g. 10\%) can partly wash out the possible existing trends. Despite this and in order to improve the statistics, in the following we use the top and lower $10\%$ of our density distribution to segregate the different environments.

\begin{table*}
    \begin{tabular}{ccccc}
      \hline
	  & $\overline{R}_{under 5\%}/\overline{R}_{over 5\%}$ & $\overline{R}_{field}/\overline{R}_{clust}$ & $\sigma_{under 5\%}/\sigma_{over  5\%}$ & $\sigma_{field}/\sigma_{clust}$\\
      \hline
	\textbf{Late-type}  &	$1.07\pm0.01$  	  &	 $1.078\pm0.006$	  & $1.01\pm0.03$	&	$1.002\pm0.008$ \\
	\textbf{Early-type} &	$1.021\pm0.006$   &      $1.040\pm0.008$	  & $1.05\pm0.01$	&   	$1.030\pm0.009$ \\
        \end{tabular}
   \caption{Same as Table \ref{table:mean_results} but using the densest $5\%$ and less dense $5\%$ of the density distribution to characterize the high and low-density environments respectively. The estimations for field and cluster galaxies are shown for comparison.}
     \label{table:mean_results5}
\end{table*}

\subsection{Comparison with previous works: the global stellar mass-size relation}
\label{sec:comparison_shen}

\begin{figure*}
  \includegraphics[width=168mm]{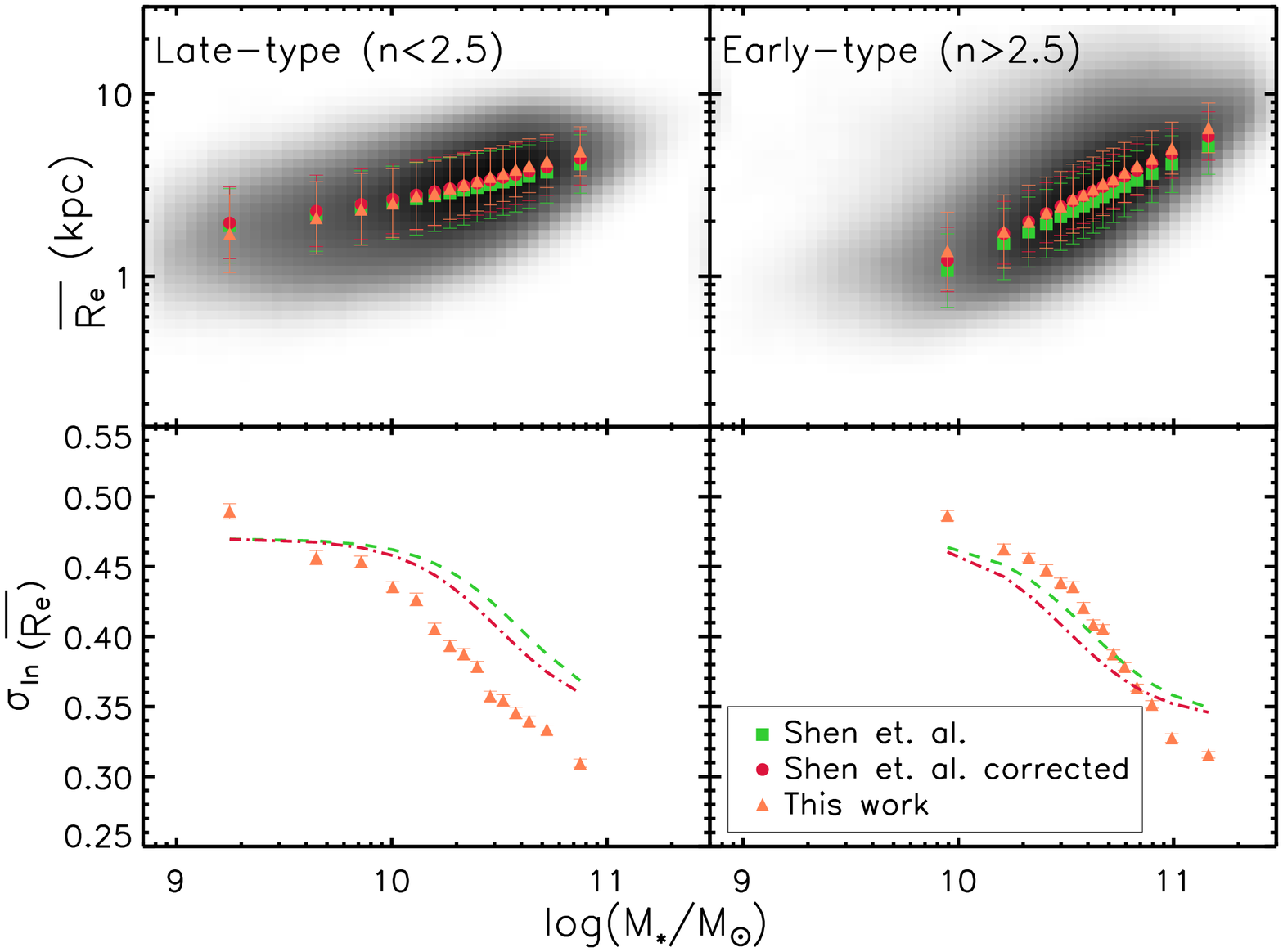}
 \caption{Comparison between the global stellar mass-size relations by Shen at al (2003) and this work. The upper panels show the stellar mass-size relation obtained in this work (orange filled triangles) and that obtained by Shen et. al. for late and early-type galaxies. Error bars represent the dispersion in the relation. The lower panels show the scatter for both morphologies. Filled orange dots represent the results of this work and the dashed green and red lines those from Shen et al (2003) and Shen et al (2003) corrected as explained in the text.}
  \label{fig:shen}
\end{figure*}

\cite{shen2003} obtained the stellar mass-size relation in the nearby Universe for both late and early-type galaxies but without segregating the galaxies according to their surrounding density. They used a different size and mass estimation, so it is worth comparing whether our results are in agreement with theirs. Figure \ref{fig:shen} shows the overall stellar mass-size relation obtained in our paper and that obtained by \cite{shen2003} as well as the dispersion of that distribution for both works.

Figure \ref{fig:shen} reveals some discrepancies between the two works in the mean sizes and the scatter. At a fixed stellar mass, the galaxies in our work are larger than those of \citet{shen2003}. This difference grows gradually for late-type galaxies as we move towards higher masses, but it is constant in the case of early-type galaxies. Also, the scatter in the stellar mass-size relations presents some differences, especially for the high mass end of both morphologies ($M_*>10^{10}M_{\sun}$ for late-type galaxies and $M_*>6\times10^{10}M_{\sun}$ for early-type galaxies) where the differences in the scatter are $\sim10\%$.

These differences between the two works can be due to several reasons. Firstly, \cite{shen2003} used stellar masses from \cite{kauffmann2003}, whereas our stellar masses are extracted using \textsc{kcorrect v4\_1\_4} (for more detail on the process see \citealt{blantonKcorrect}). According to \citet{blantonKcorrect}, \citet{kauffmann2003} overestimate the stellar masses, with differences about $\sim0.1$ dex for stellar masses $M_*>4\times10^{9}M_{\sun}$. This means that our masses should be a factor $1.26$ smaller than those used in \citet{shen2003}. Figure \ref{fig:shen} (upper panels) shows the \cite{shen2003} relation using this conversion factor of $1.26$ between \citet{blantonKcorrect} and \citet{kauffmann2003} masses. It is clear from the plot that this correction makes both stellar mass-size relations practically indistinguishable from each other, except in the higher bins of mass. Nevertheless, the scatter of the relation still shows some differences, especially at high mass. In what follows, we suggest a 
potential explanation for this fact.

To explain the differences in the scatter between the \cite{shen2003} sample and ours, we must take into account the size evolution of the galaxies. Since their sample spans over a wider range of redshift than ours, \cite{shen2003} include galaxies with redshifts up to $z\sim0.3$ with a median redshift of $z\sim0.1$. It has been previously found (\citealt{trujillo2004,daddi2005, trujillo2006, trujillo2007} and references therein) that galaxies have suffered a strong size evolution since $z\sim2$, being larger as we move towards the present-day Universe. The wider range of redshift considered in \cite{shen2003} potentially includes galaxies from a wider range of epochs in their stellar mass-size relation. This produces a bias towards more compact objects at a fixed mass, increasing the resulting scatter and introducing and offset respect to our stellar mass-size relation. We reduce this issue by restricting our sample to galaxies within the redshift range $z=0.005-0.12$. In their sample, the larger redshift 
range could translate in a wider distribution of sizes at a given stellar mass. It is very important to note that higher redshifts (and thus size evolution) contribute especially in their high mass end of the distributions, since lower stellar mass galaxies at higher redshift are not detected (see Figure \ref{fig:completeness}). If our explanation is correct, we should expect that the scatter of the \cite{shen2003} relations were larger than ours at the highest mass bins. This is in fact the case.

When we take these factors into account, we can see that our results are in better qualitative agreement with those of \cite{shen2003}. It is also worth mentioning that our sample is a factor of 2 larger than the sample used in \citet{shen2003} and with a more restricted redshift range, providing better statistics to our work in our redshift range.

\subsection{Comparison with previous works: Mean sizes of the galaxies as a function of the environment}
\label{sec:comparison}

Low redshift ($z<0.3$) studies have repeatedly reported larger sizes for late-type galaxies residing in low-density environments compared to high-density regions. However, no statistically significant trend for early-type galaxies has been reported (e.g. \citealt{maltby2010,valentinuzzi2010,fernandez2013,poggianti2013}). In this work, we have showed that both late and early-type galaxies present larger sizes in the field (an average value of $\sim7.5\%$ and $\sim3.5\%$ respectively) compared to high-density environments. Despite previous works are qualitatively in line with the results of our work, we find interesting to asses here the quantitative differences between them.

\subsubsection{The Valentinuzzi et al (2010) and Poggianti et al (2013) samples}

\cite{valentinuzzi2010} and \cite{poggianti2013} used the WINGS survey to obtain a sample of $\sim600$ cluster galaxies in the redshift range $0.04<z<0.07$. Both works compare the sizes of this sample of cluster galaxies with sizes in the general field at a fixed stellar mass using a \cite{kroupa2001} IMF. \cite{valentinuzzi2010} use the local relation found by \cite{shen2003} (discussed in Section \ref{sec:comparison_shen}) as representative of field galaxies in a stellar mass range similar to that used in our work ($M_*<4\times10^{11}M_{\sun}$). They find an offset between the general relation in the field and that followed by their cluster galaxies, the later showing smaller radii at fixed stellar mass. They claim that this difference can be due to the offset between stellar masses in their samples, drawn from different catalogues. On the other hand, \cite{poggianti2013} use a similar cluster sample for galaxies with masses $M_*<3\times10^{11}M_{\sun}$ and compare it with both a sample extracted from 
PM2GC \citep{calvi2011} and the local relation by \cite{shen2003}, also finding an offset of about 1$-\sigma$ between both relations. 

\begin{figure*}
  \includegraphics[width=168mm]{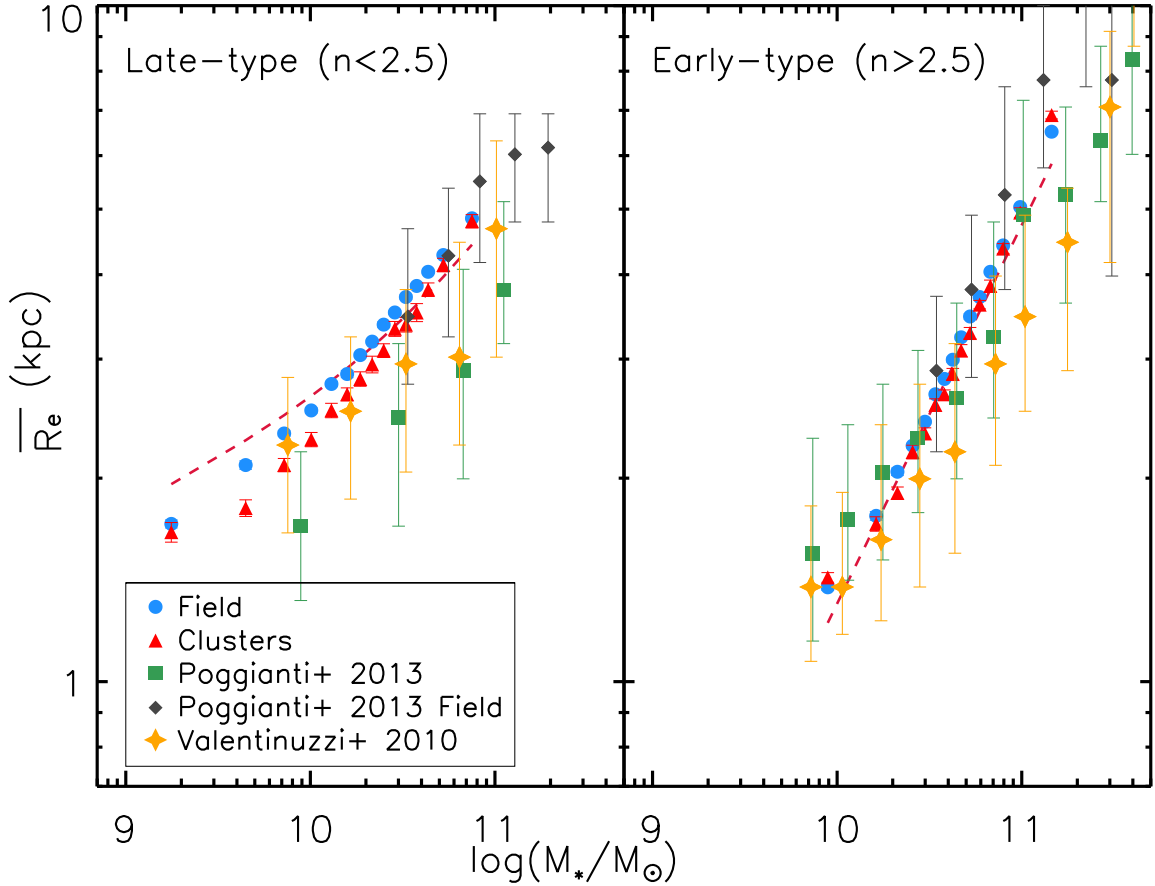}
 \caption{The stellar mass-size relation for field and cluster galaxies. The red filled triangles and blue filled dots represent the results obtained in this work for clusters and field galaxies. The red dashed line is the local relation extracted from Shen et al (2003) and corrected as explained in Section \ref{sec:comparison_shen}. Green squares and orange stars represent the results from Poggianti et al (2013) and Valentinuzzi et al (2010) with error bars showing the lower and upper quartiles of the distribution.}
  \label{fig:compare}
\end{figure*}

Figure \ref{fig:compare} shows the results obtained by \cite{valentinuzzi2010} and \cite{poggianti2013}, rescaled to a \cite{chabrier2003} IMF, together with the stellar mass-size relation obtained in this work for cluster and field galaxies. Despite using almost the same sample as \cite{valentinuzzi2010}, cluster early-type galaxies from \cite{poggianti2013} seem larger at a fixed stellar mass than those from \cite{valentinuzzi2010}. The opposite trend is found for late-type objects. This discrepancy could be due either to the different morphological segregation between both works or the filter in which these structural parameters were measured. While \cite{poggianti2013} use the S\'ersic index for segregating disk-like and spheroid-like galaxies ($n<2.5$ and $n>2.5$ respectively) in the \textit{B}-band, \cite{valentinuzzi2010} use \textsc{MORPHOT} \citep{fasano2012} in \textit{V}-band to classify galaxies in two groups: S0 and ellipticals in one of them and galaxies later than S0 in the other. Using different bands for measuring structural parameters can slightly affect the resulting sizes of the galaxies. As a rough approximation, we measure that the average difference in the effective radius measured in \textit{z}-band (used in \citealt{shen2003}) and \textit{g}-band is $<R_{eff,z}-R_{eff,g}>\sim0.3$ kpc. 

Despite these differences, for both samples the cluster galaxies are $\sim30\%$ smaller than the galaxies in the field. Although this is in qualitative agreement with our work, we want to stress that differences among environments are less evident in our data. It is worth noting that we consider spheres of 2 Mpc radius, while the mean aperture in \cite{valentinuzzi2010} and \cite{poggianti2013} is of about $\sim1.3$ Mpc (see \citealt{varela2009} and \citealt{cava2009}). To test if the aperture can be responsible for these differences, Table \ref{table:aperture} presents the results using different apertures. Using an aperture similar to that used in \cite{valentinuzzi2010} and \cite{poggianti2013} does not significantly change our results, though it increases the errors as expected due to the lower statistics in each bin.

\begin{table*}
    \begin{tabular}{ccccc}
      \hline
	  Aperture &  \multicolumn{2}{c}{Late-type} & \multicolumn{2}{c}{Early-type}\\
	  & $\overline{R}_{field}/\overline{R}_{clust}$  & $\sigma_{field}/\sigma_{clust}$ & $\overline{R}_{field}/\overline{R}_{clust}$  & $\sigma_{field}/\sigma_{clust}$\\
      \hline
	0.5 Mpc  &	$1.21\pm0.03$  	    &	   $1.04\pm0.05$ 	  & $1.05\pm0.02$	&	$1.08\pm0.03$ \\
	1.0 Mpc  &	$1.13\pm0.02$       &      $1.05\pm0.02$	  & $1.05\pm0.01$	&   	$1.08\pm0.02$ \\
	1.5 Mpc  &	$1.10\pm0.01$       &      $1.05\pm0.02$	  & $1.046\pm0.008$	&   	$1.04\pm0.02$ \\
	2.0 Mpc  &	$1.078\pm0.006$     &      $1.002\pm0.008$	  & $1.040\pm0.008$	&   	$1.030\pm0.009$ \\
        \end{tabular}
   \caption{Comparison of the results obtained using different radial apertures for defining the environmental density around our galaxies. The table shows the results for late and early-type galaxies in clusters and in the field.}
     \label{table:aperture}
\end{table*}

Table \ref{table:aperture} also shows how the difference between the sizes in different environments becomes larger as we reduce the searching aperture for defining the environmental density. This result indicates that the effect of the environment becomes more intense at the innermost region of the clusters. These are the most extreme environments where the tidal effects of the other cluster members as well as the gas pressure of the cluster can play a larger role. This is particularly important for the late-type galaxies whereas the early-type galaxies seem to be less affected.

\subsubsection{The Fernandez-Lorenzo et al (2013) sample}

\cite{fernandez2013} found qualitatively similar results to \cite{valentinuzzi2010,poggianti2013} and those in our work in a sample of 452 isolated galaxies with masses $M_*<3\times10^{10}M_{\sun}$ drawn from the AMIGA project \citep{2005verdes-montenegro}. As in our work, stellar masses were obtained using \cite{chabrier2003} IMF. The circularized radii for the galaxies in their sample were obtained using a S\'ersic profile. Again, the local stellar mass-size relation by \cite{shen2003} was used as a representative for the general field population. The isolated galaxies sample was segregated according to their S\'ersic indexes ($n<2.5$ for late-type and $n>2.5$ for early-type galaxies) and fitted with the same function as in \cite{shen2003} but leaving the zero point of the relation as a free parameter of the fit. By comparing the obtained zero point with that from \cite{shen2003}, they find that isolated early-type galaxies show the same size at a fixed stellar mass than the overall distribution of galaxies. Late-type galaxies, conversely, are $\sim17\%$ larger when 
isolated. This result, although qualitatively in line with our work, is not straightforward to compare with the results shown in our paper. As explained in Section \ref{sec:ndensity}, fixed aperture methods are not especially sensitive in low-density regions. This means that our sample of low-density galaxies can include less isolated objects than those considered in the AMIGA sample, leading to a difficulty when comparing the two samples directly. Although it is beyond the scope of this paper, it would be very interesting to asses an analogous study with environment estimators more suited to low-density regions such as that of \cite{fernandez2013}.

\subsubsection{The Maltby et al (2010) sample}

Finally, \cite{maltby2010} carried out a similar study to ours in the nearby Universe using 734 galaxies belonging to the A901/902 supercluster at $z\sim0.17$. Field galaxies were extracted from the same field of $5\times5$ Mpc from the STAGES survey but using redshift slides placed before and after the supercluster structure. For the lower mass end in their sample ($M_*<10^{10}M_{\sun}$, computed using \citealt{kroupa2001} IMF), visually classified cluster spirals show $\sim17\%$ larger semi-major axis in the \textit{V}-band than cluster spirals, in agreement with our results for cluster galaxies within the error bars (see Figure \ref{fig:diffR-clusters}). This trend is enhanced when only core spirals are selected in line with our results shown in Table \ref{table:aperture}. In contrast, their elliptical and S0 galaxies do not show any difference.

It is worth noting that our work present several advantages compared with previous works. On one hand, our sample is homogeneous and drawn from the same catalogue, which makes it self-consistent. Potential problems of size measurement in our catalogue due to sky oversubstraction \citep{guo2009} affect our sample homogeneously and, henceforth, do not influence our results. This consistency avoids biases among magnitudes extracted from different sets of data which often implies different methods, instruments and tools, commonly found in the literature when comparing different environments. On the other hand, the size of our sample also provides an advantage, allowing to probe a wider range of mass and reducing the statistical errors significantly. In addition, we use two independent methods for estimating the environment where each of the galaxies in our sample reside. Both these methods reproduce the same trends with similar values and errors. Finally, our main sample has been carefully selected to avoid the 
biases towards young galaxies derived from a pure magnitude selection, as explained in Section \ref{sec:completeness}.

\subsection{The scatter of the stellar mass-size relation}

Figures \ref{fig:diffS-neigh} and \ref{fig:diffS-clusters} show the results obtained for the dispersion of the stellar mass-size relation for the samples used in this work. These numbers are in qualitative agreement with those of \cite{shen2003} (see Figure \ref{fig:shen}). However, we have measured a slightly smaller scatter, probably due to evolution effects in their sample (see Section \ref{sec:comparison_shen}).

Our analyses show that the environmental effect on the scatter of the size distribution strongly depends on the stellar mass. At a fixed stellar mass the scatter is higher for late-type galaxies inhabiting overdense regions and with masses below $2\times10^{10}M_{\sun}$. For masses greater than that, the scatter does not show any environmental effect within the error bars. For early-type galaxies, the stellar mass dependence of the scatter is weaker, but the average effect is more pronounced: early-type galaxies in clusters or overdensities are about $\sim4\%$ less scattered around the mean size than their counterparts in the field or underdensities. Again, these results are fully consistent between the two different methods used here to estimate the environment where the galaxy resides.

To the best of our knowledge no previous work have quantified the effect of the environment on the scatter of the stellar mass-size relation. This new insight of our study is only possible due to our statistically significant sample. It is worth noting that \cite{fernandez2013} qualitatively mention that isolated objects in their sample present a more tight stellar mass-size relation than their sample without environmental segregation, contrary to our results. Nevertheless, as stated in Section \ref{sec:comparison}, a comparison between their work and ours is not straightforward. Also, the smaller size of their sample could lead to an underestimation of the scatter.

\section{Discussion}
\label{sec:discussion}

Our work shows that both the average size and the scatter in the stellar mass-size distribution are larger in low-density environments that in high-density regions. Those effects are stronger in late-type galaxies than in early-type objects. Moreover, the effects increase when we explore the innermost region of the galaxy clusters. What are these trends telling us about the formation and evolution of the galaxies?

Recent semi-analytical models \citep{shankar2013,shankar2014} predict a moderate effect of the environment in the size of the galaxies in the nearby Universe. These models typically segregate the different environments based on their halo mass . According to the models of \cite{shankar2013,shankar2014}, galaxies in more massive halos have larger sizes than galaxies with similar stellar masses inhabiting less massive halos. This is due to the larger number of mergers suffered during their history \citep{shankar2013}. However, the opposite trend is found observationally: late-type galaxies are slightly more compact in clusters than in the field \citep{maltby2010, valentinuzzi2010,fernandez2013,poggianti2013}, and elliptical galaxies do not show any dependence on the environment \citep{maltby2010,huertas2013a,huertas2013b}. Our results  support this view: galaxies are larger in the field, independently of their morphology, although early-type galaxies seem to be less sensitive to the environment than disc galaxies. 

Interestingly, as we move towards higher redshifts, objects located in high-density regions start to show larger sizes ($25\%$ to $50\%$ larger) than those placed in low-density environments. This is shown by different works both at intermediate redshift $z\sim1$: \cite{cooper2012,delaye2013} and \cite{papovich2012} (but see also \citealt{huertas2013a,rettura2010} and \citealt{raichoor2012} for different conclusions); and also at higher redshifts ($z\sim2$, \citealt{strazzullo2013}). \cite{lani2013} studied the environmental dependence of the size for early-type galaxies in two redshift ranges ($0.5<z<1$ and $1<z<2$). At high redshift, they find larger early-type galaxies in dense environments than in underdense environments, consistent with the previously mentioned studies, but the trend weakens for their lower redshift range.

Supporting these observations, \cite{maulbetsch2007} found in N-body simulations that for halos in lower density regions the average present-day aggregation rate is 4-5 times higher than for halos in high-density regions. As we move towards higher redshift, their simulations show that this difference becomes smaller until $z\geqslant1$, when the trend is reversed and higher density environments present higher accretion rates than lower density environments. This different mass accretion rate in different environments can explain why results at high redshift differ from those in the nearby Universe.

All the above observational and theoretical evidence points towards a scenario in which at earlier epochs ($z>1$) galaxies in high-density regions could have undergone a faster growth than the galaxies in less dense regions. However, in the last 7 Gyr this growth may have slowed down in clusters, while maintained, even increased, in the field, allowing low-density galaxies to reach similar sizes than the ones we observe in the cluster nowadays.

When comparing observational studies with simulations, it is important to take into account the role that errors can play in the results. \cite{huertas2013b} and \cite{shankar2014} investigated the possible effect that errors in masses and effective radii can have on measuring environmental effects on the stellar mass-size distribution in the nearby Universe. Both studies indicate that the lack of environmental effects reported in previous works in early-type galaxies could be due to observational errors in measuring both the stellar mass and the effective radius. \cite{huertas2013b} quantify this effect claiming that differences in sizes of $\sim40\%$ at a fixed stellar mass would not be detected with significance. Although quantifying the effect of individual galaxy errors is beyond the scope of this work, it is worth mentioning that our approach should reduce the influence of observational errors and show any possible trend on the size of the galaxies with the environment due to the increase in the size of the sample. Nevertheless, and despite the improvement in the statistical significance of our work compared to previous samples, the effect of the environment we find is very mild and the opposite to that predicted by \cite{shankar2014}.

In relation to the scatter in the stellar mass-size plane, it is worth noting that the models predict a scatter in sizes at a fixed stellar mass in disagreement with the tight relations found in the data \citep{nipoti2009,nair2011,shankar2013}. The observed low values for the scatter can only be reproduced by the models requiring a fine tuning in the input scaling relations of the progenitors \citep{shen2003,shankar2014}, forcing them to follow also tight scaling relations. \cite{nipoti2012} points out that even without considering some sources of scatter (such as that associated to the intrinsic scatter of stellar to halo mass relation), the scatter found in simulations is larger than that obtained observationally.

It is important to take into account that the value for the scatter obtained in our work is an upper limit of the intrinsic scatter of the stellar mass-size distribution. The observed scatter includes two different contributions: the intrinsic scatter of the stellar mass-size distribution, due to the physical processes taking place in each galaxy, and the scatter induced because of the errors in stellar mass and size measurements. To have a hint on the scatter due to observational errors, we have taken the average stellar mass-size relations obtained in the previous sections and blurred them by generating random points according to the typical error values both in stellar mass and in size. Then, we evaluate the mean size and the scatter of the distribution using our likelihood estimators. Our tests show that scatter due to realistic observational errors ($\frac{\delta \log M_{*}}{\log M_{*}}=0.2$;  $\frac{\delta \overline{R}_e}{\overline{R}_e}=0.15$; \citealt{blantonKcorrect,blantonNYU}) is around $<\sigma_{\ln R}>\sim0.20$ for late-type galaxies and $<\sigma_{\ln R}>\sim0.27$ for early-type galaxies. Consequently, the intrinsic scatter of the distribution would be $<\sigma_{\ln R}>\sim0.33$ in the case of late-type galaxies and $<\sigma_{\ln R}>\sim0.37$ for early-type galaxies. These small scatter values would make the discrepancy with the theoretical values even larger. Appendix \ref{sec:appendixB} details the tests carried out to estimate the above values.

We find that the scatter in sizes at a fixed stellar mass is slightly smaller in clusters or overdense regions. \cite{nair2011} found a similar result when studying the luminosity-size relation. This result could be connected with a faster evolution and formation of the galaxies in clusters, thus involving a more homogeneous population of the galaxies during the merging processes at high redshift. If our suggestion is correct, the distribution of cluster galaxies in the stellar mass-size plane would reflect a more primordial distribution of sizes, maybe related to progenitors with tighter scaling relations, as required in the models. On the contrary, galaxies in the field could have longer evolutionary time-scales and the objects nearby them would have more time to evolve before they merge into the central galaxy. This could broaden the characteristics of the final objects and therefore broaden the dispersion of the distribution in the stellar mass-size plane.

In general, the influence of the environment on the distribution of sizes at fixed stellar mass is stellar mass-dependent and seems more pronounced on low to intermediate stellar masses ($M_*<2\times10^{10}M_{\sun}$). Unfortunately, these masses are harder to study with high statistics and only at very low redshifts can we probe mass-complete samples. Thus, a greater effort to explore this range of stellar masses at progressively higher redshifts has to be made in order to fully characterize the effect of the environment on the size of the galaxies. Exploring the stellar mass-size relation and its scatter at higher redshift can also help to disentangle the effect that  merging can have on the growing of galaxies and test other proposed mechanisms for the evolution of the stellar mass-size relation such as the arrival of newcomers \citep[see e.g.][]{vanderwel2005,carollo2013}. We also want to stress that comparing different samples from different works is often not straightforward. Factors such as the IMF used, the methods for measuring structural parameters, band in which parameters are measured or the inhomogeneity of the main samples can cover subtle effects or even artificially reinforce them.

At higher redshifts the environment where the galaxies grow leaves an imprint in the size of the galaxies with galaxies in clusters being $30-50\%$ larger than in the field (see \citealt{delaye2013, lani2013,papovich2012}, but see also \citealt{huertas2013a, rettura2010} and \citealt{raichoor2012} for a different view). In the nearby universe, however, the environment plays a minor role, which many current galaxy formation models have failed to predict.

We want to end this section by pointing out the profound mystery that our data present. As we have mentioned above, both the numerical simulations and the observations at high redshift show a significantly different evolutionary speed in clusters than in the field. Also, the characteristics of the progenitors should be different depending on the environment. Despite that, in the present-day Universe, the environmental differences in both the mean sizes and the scatter of the galaxies are small. Why have the galaxies today reached such similar sizes both in the clusters and in the field, when they have followed strikingly different growth histories? This is still an open question.

\section{Summary}
\label{sec:summary}

In this work we have studied the distribution of the galaxies in the stellar mass-size plane for a mass-complete sample of nearby ($z<0.12$) 232000 objects. We use the NYU-VAGC catalogue to obtain a wide range of galaxies with different stellar masses located in different environments. At fixed stellar mass, we have computed the mean size and the scatter in the distribution using a maximum likelihood method. This calculation has been done using two different estimators of the environment. On one hand, for each galaxy in our sample, we have computed the number of galaxies with masses above $4\times10^{10}M_{\sun}$ within a sphere of 2 Mpc radius and taken the total mass inside this sphere as an indicator of the surrounding density. We take the 10$\%$ of galaxies having the lower and the 10$\%$ having the larger density values as representative samples of underdense and overdense zones. On the other hand, we have compiled a large number of galaxy clusters within our redshift range and compared the 
galaxies within 2 Mpc from the centre of the clusters with the galaxies in the field. In both approaches, we have segregated our samples according to their morphology based on the S\'ersic index.

Regarding the size of the galaxies at a fixed stellar mass, we find that the galaxies are slightly larger ($\sim7.5\%$ for late-type galaxies and $\sim3.5\%$ for early-type galaxies) in less dense environments than in high-density regions. Our result is statistically significant  for both late and early-type galaxies and widely consistent among the two methods used to characterize the environment. Qualitative similar results have been previously found for late-type galaxies in several works in a comparable redshift range (\citealt{poggianti2013, fernandez2013} and \citealt{maltby2010}).

Our work, however, is the first to claim that, on average, early-type galaxies in low-density environments are about $\sim3.5\%$ larger than in clusters with high statistical significance ($>4\sigma$). This result have not been found in previous works \citep{maltby2010,huertas2013b,fernandez2013}, probably because the effect is subtle and requires a large sample to unveil this environmental effect. \cite{poggianti2013} found qualitatively the same trend but with low significance due to the lower number of galaxies in their sample (a factor of $\sim40$ smaller).

The scatter of the stellar mass-size distribution has also been explored in this work. Our results show that the environment does not change significantly the scatter of the overall distribution. Nevertheless, there is a hint of the distribution having less scatter in high-density regions, especially for early-type galaxies and for late-type galaxies with masses below $2\times10^{10}M_{\sun}$. We also confirm previous observational values for the scatter of the overall stellar mass-size relation such as those from \cite{shen2003} and \cite{nair2011}. 

Taken together, our results point towards an earlier evolution of galaxies in clusters from a more constrained family of progenitors. This evolution may have slowed down in the past few Gyr allowing objects in less dense environments to reach similar sizes to those located in high-density regions. This different evolutionary speed between galaxies in clusters and in the field could have lead to the weak correlation between environment and mean sizes at a fixed stellar mass observed in the present-day Universe.

\section*{Acknowledgments}

We wish to thank the anonymous referee for constructive comments which helped to improve the paper. We thank Martin Stringer for interesting comments. We also want to thank Luis Peralta de Arriba and Andr\'es del Pino for useful suggestions that have helped to improve the paper. This work has been supported by the Programa Nacional de Astronom\'ia y Astrof\'isica of the Spanish Ministry of Economy and competitiveness under grants SEV-2011-0187 and AYA2010-21322-C03-02.

Funding for the SDSS and SDSS-II has been provided by the Alfred P. Sloan Foundation, the Participating Institutions, the National Science Foundation, the U.S. Department of Energy, the National Aeronautics and Space Administration, the Japanese Monbukagakusho, the Max Planck Society, and the Higher Education Funding Council for England. The SDSS Web Site is http://www.sdss.org/.

The SDSS is managed by the Astrophysical Research Consortium for the Participating Institutions. The Participating Institutions are the American Museum of Natural History, Astrophysical Institute Potsdam, University of Basel, University of Cambridge, Case Western Reserve University, University of Chicago, Drexel University, Fermilab, the Institute for Advanced Study, the Japan Participation Group, Johns Hopkins University, the Joint Institute for Nuclear Astrophysics, the Kavli Institute for Particle Astrophysics and Cosmology, the Korean Scientist Group, the Chinese Academy of Sciences (LAMOST), Los Alamos National Laboratory, the Max-Planck-Institute for Astronomy (MPIA), the Max-Planck-Institute for Astrophysics (MPA), New Mexico State University, Ohio State University, University of Pittsburgh, University of Portsmouth, Princeton University, the United States Naval Observatory, and the University of Washington.

\footnotesize{

\bibliographystyle{mn2e_fixed}
\bibliography{bibliography}

\begin{thebibliography}{81}
\expandafter\ifx\csname natexlab\endcsname\relax\def\natexlab#1{#1}\fi

\bibitem[{{Abazajian} {et~al}\mbox{.}(2009){Abazajian}, {Adelman-McCarthy},
  {Ag{\"u}eros}, {Allam}, {Allende Prieto}, {An}, {Anderson}, {Anderson},
  {Annis}, {Bahcall}, \& et~al.}]{sdss-dr7}
{Abazajian} K.~N. {et~al.}, 2009, \apjs, 182, 543

\bibitem[{{Abell}, {Corwin} \& {Olowin}(1989){Abell}, {Corwin}, \&
  {Olowin}}]{abell1989}
{Abell} G.~O., {Corwin}, Jr. H.~G., {Olowin} R.~P., 1989, \apjs, 70, 1

\bibitem[{{Andredakis}, {Peletier} \& {Balcells}(1995){Andredakis}, {Peletier},
  \& {Balcells}}]{andredakis1995}
{Andredakis} Y.~C., {Peletier} R.~F., {Balcells} M., 1995, \mnras, 275, 874

\bibitem[{{Barden} {et~al}\mbox{.}(2005){Barden}, {Rix}, {Somerville}, {Bell},
  {H{\"a}u{\ss}ler}, {Peng}, {Borch}, {Beckwith}, {Caldwell}, {Heymans},
  {Jahnke}, {Jogee}, {McIntosh}, {Meisenheimer}, {S{\'a}nchez}, {Wisotzki}, \&
  {Wolf}}]{barden2005}
{Barden} M. {et~al.}, 2005, \apj, 635, 959

\bibitem[{{Bezanson} {et~al}\mbox{.}(2013){Bezanson}, {van Dokkum}, {van de
  Sande}, {Franx}, {Leja}, \& {Kriek}}]{bezanson2013}
{Bezanson} R., {van Dokkum} P.~G., {van de Sande} J., {Franx} M., {Leja} J.,
  {Kriek} M., 2013, \apjl, 779, L21

\bibitem[{{Blanton} {et~al}\mbox{.}(2005{\natexlab{a}}){Blanton}, {Eisenstein},
  {Hogg}, {Schlegel}, \& {Brinkmann}}]{blanton2005}
{Blanton} M.~R., {Eisenstein} D., {Hogg} D.~W., {Schlegel} D.~J., {Brinkmann}
  J., 2005{\natexlab{a}}, \apj, 629, 143

\bibitem[{{Blanton} \& {Roweis}(2007)}]{blantonKcorrect}
{Blanton} M.~R., {Roweis} S., 2007, \aj, 133, 734

\bibitem[{{Blanton} {et~al}\mbox{.}(2005{\natexlab{b}}){Blanton}, {Schlegel},
  {Strauss}, {Brinkmann}, {Finkbeiner}, {Fukugita}, {Gunn}, {Hogg},
  {Ivezi{\'c}}, {Knapp}, {Lupton}, {Munn}, {Schneider}, {Tegmark}, \&
  {Zehavi}}]{blantonNYU}
{Blanton} M.~R. {et~al.}, 2005{\natexlab{b}}, \aj, 129, 2562

\bibitem[{{Bruzual} \& {Charlot}(2003)}]{Bruzual&Charlot}
{Bruzual} G., {Charlot} S., 2003, \mnras, 344, 1000

\bibitem[{{Buitrago} {et~al}\mbox{.}(2008){Buitrago}, {Trujillo}, {Conselice},
  {Bouwens}, {Dickinson}, \& {Yan}}]{buitrago2008}
{Buitrago} F., {Trujillo} I., {Conselice} C.~J., {Bouwens} R.~J., {Dickinson}
  M., {Yan} H., 2008, \apjl, 687, L61

\bibitem[{{Buitrago} {et~al}\mbox{.}(2013){Buitrago}, {Trujillo}, {Conselice},
  \& {H{\"a}u{\ss}ler}}]{buitrago2013}
{Buitrago} F., {Trujillo} I., {Conselice} C.~J., {H{\"a}u{\ss}ler} B., 2013,
  \mnras, 428, 1460

\bibitem[{{Calvi}, {Poggianti} \& {Vulcani}(2011){Calvi}, {Poggianti}, \&
  {Vulcani}}]{calvi2011}
{Calvi} R., {Poggianti} B.~M., {Vulcani} B., 2011, \mnras, 416, 727

\bibitem[{{Cappellari}(2013)}]{cappellari2013}
{Cappellari} M., 2013, \apjl, 778, L2

\bibitem[{{Carollo} {et~al}\mbox{.}(2013){Carollo}, {Bschorr}, {Renzini},
  {Lilly}, {Capak}, {Cibinel}, {Ilbert}, {Onodera}, {Scoville}, {Cameron},
  {Mobasher}, {Sanders}, \& {Taniguchi}}]{carollo2013}
{Carollo} C.~M. {et~al.}, 2013, \apj, 773, 112

\bibitem[{{Cassata} {et~al}\mbox{.}(2011){Cassata}, {Giavalisco}, {Guo},
  {Renzini}, {Ferguson}, {Koekemoer}, {Salimbeni}, {Scarlata}, {Grogin},
  {Conselice}, {Dahlen}, {Lotz}, {Dickinson}, \& {Lin}}]{cassata2011}
{Cassata} P. {et~al.}, 2011, \apj, 743, 96

\bibitem[{{Cassata} {et~al}\mbox{.}(2013){Cassata}, {Giavalisco}, {Williams},
  {Guo}, {Lee}, {Renzini}, {Ferguson}, {Faber}, {Barro}, {McIntosh}, {Lu},
  {Bell}, {Koo}, {Papovich}, {Ryan}, {Conselice}, {Grogin}, {Koekemoer}, \&
  {Hathi}}]{cassata2013}
{Cassata} P. {et~al.}, 2013, \apj, 775, 106

\bibitem[{{Cava} {et~al}\mbox{.}(2009){Cava}, {Bettoni}, {Poggianti}, {Couch},
  {Moles}, {Varela}, {Biviano}, {D'Onofrio}, {Dressler}, {Fasano}, {Fritz},
  {Kj{\ae}rgaard}, {Ramella}, \& {Valentinuzzi}}]{cava2009}
{Cava} A. {et~al.}, 2009, \aap, 495, 707

\bibitem[{{Chabrier}(2003)}]{chabrier2003}
{Chabrier} G., 2003, \pasp, 115, 763

\bibitem[{{Ciotti}, {Lanzoni} \& {Volonteri}(2007){Ciotti}, {Lanzoni}, \&
  {Volonteri}}]{ciotti2007}
{Ciotti} L., {Lanzoni} B., {Volonteri} M., 2007, \apj, 658, 65

\bibitem[{{Cooper} {et~al}\mbox{.}(2012){Cooper}, {Griffith}, {Newman}, {Coil},
  {Davis}, {Dutton}, {Faber}, {Guhathakurta}, {Koo}, {Lotz}, {Weiner},
  {Willmer}, \& {Yan}}]{cooper2012}
{Cooper} M.~C. {et~al.}, 2012, \mnras, 419, 3018

\bibitem[{{Cooper} {et~al}\mbox{.}(2005){Cooper}, {Newman}, {Madgwick},
  {Gerke}, {Yan}, \& {Davis}}]{cooper2005}
{Cooper} M.~C., {Newman} J.~A., {Madgwick} D.~S., {Gerke} B.~F., {Yan} R.,
  {Davis} M., 2005, \apj, 634, 833

\bibitem[{{Daddi} {et~al}\mbox{.}(2005){Daddi}, {Renzini}, {Pirzkal},
  {Cimatti}, {Malhotra}, {Stiavelli}, {Xu}, {Pasquali}, {Rhoads}, {Brusa}, {di
  Serego Alighieri}, {Ferguson}, {Koekemoer}, {Moustakas}, {Panagia}, \&
  {Windhorst}}]{daddi2005}
{Daddi} E. {et~al.}, 2005, \apj, 626, 680

\bibitem[{{Delaye} {et~al}\mbox{.}(2013){Delaye}, {Huertas-Company}, {Mei},
  {Lidman}, {Licitra}, {Newman}, {Raichoor}, {Shankar}, {Barrientos},
  {Bernardi}, {Cerulo}, {Couch}, {Demarco}, {Mu{\~n}oz}, {Sanchez-Janssen}, \&
  {Tanaka}}]{delaye2013}
{Delaye} L. {et~al.}, 2013, ArXiv, 1307.0003

\bibitem[{{Einasto} {et~al}\mbox{.}(2012){Einasto}, {Liivam{\"a}gi}, {Tempel},
  {Saar}, {Vennik}, {Nurmi}, {Gramann}, {Einasto}, {Tago}, {Hein{\"a}m{\"a}ki},
  {Ahvensalmi}, \& {Mart{\'{\i}}nez}}]{einasto2012}
{Einasto} M. {et~al.}, 2012, \aap, 542, A36

\bibitem[{{Fasano} {et~al}\mbox{.}(2012){Fasano}, {Vanzella}, {Dressler},
  {Poggianti}, {Moles}, {Bettoni}, {Valentinuzzi}, {Moretti}, {D'Onofrio},
  {Varela}, {Couch}, {Kj{\ae}rgaard}, {Fritz}, {Omizzolo}, \&
  {Cava}}]{fasano2012}
{Fasano} G. {et~al.}, 2012, \mnras, 420, 926

\bibitem[{{Fern{\'a}ndez Lorenzo} {et~al}\mbox{.}(2013){Fern{\'a}ndez Lorenzo},
  {Sulentic}, {Verdes-Montenegro}, \& {Argudo-Fern{\'a}ndez}}]{fernandez2013}
{Fern{\'a}ndez Lorenzo} M., {Sulentic} J., {Verdes-Montenegro} L.,
  {Argudo-Fern{\'a}ndez} M., 2013, \mnras, 434, 325

\bibitem[{{Gao}, {Springel} \& {White}(2005){Gao}, {Springel}, \&
  {White}}]{gao2005}
{Gao} L., {Springel} V., {White} S.~D.~M., 2005, \mnras, 363, L66

\bibitem[{{Guo} {et~al}\mbox{.}(2009){Guo}, {McIntosh}, {Mo}, {Katz}, {van den
  Bosch}, {Weinberg}, {Weinmann}, {Pasquali}, \& {Yang}}]{guo2009}
{Guo} Y. {et~al.}, 2009, \mnras, 398, 1129

\bibitem[{{Haas}, {Schaye} \& {Jeeson-Daniel}(2012){Haas}, {Schaye}, \&
  {Jeeson-Daniel}}]{haas2012}
{Haas} M.~R., {Schaye} J., {Jeeson-Daniel} A., 2012, \mnras, 419, 2133

\bibitem[{{Hao} {et~al}\mbox{.}(2010){Hao}, {McKay}, {Koester}, {Rykoff},
  {Rozo}, {Annis}, {Wechsler}, {Evrard}, {Siegel}, {Becker}, {Busha}, {Gerdes},
  {Johnston}, \& {Sheldon}}]{hao2010}
{Hao} J. {et~al.}, 2010, \apjs, 191, 254

\bibitem[{{Harker} {et~al}\mbox{.}(2006){Harker}, {Cole}, {Helly}, {Frenk}, \&
  {Jenkins}}]{harker2006}
{Harker} G., {Cole} S., {Helly} J., {Frenk} C., {Jenkins} A., 2006, \mnras,
  367, 1039

\bibitem[{{Huertas-Company}
  {et~al}\mbox{.}(2013{\natexlab{a}}){Huertas-Company}, {Mei}, {Shankar},
  {Delaye}, {Raichoor}, {Covone}, {Finoguenov}, {Kneib}, {Le}, \&
  {Povic}}]{huertas2013a}
{Huertas-Company} M. {et~al.}, 2013{\natexlab{a}}, \mnras, 428, 1715

\bibitem[{{Huertas-Company}
  {et~al}\mbox{.}(2013{\natexlab{b}}){Huertas-Company}, {Shankar}, {Mei},
  {Bernardi}, {Aguerri}, {Meert}, \& {Vikram}}]{huertas2013b}
{Huertas-Company} M., {Shankar} F., {Mei} S., {Bernardi} M., {Aguerri}
  J.~A.~L., {Meert} A., {Vikram} V., 2013{\natexlab{b}}, \apj, 779, 29

\bibitem[{{Kauffmann} {et~al}\mbox{.}(2003){Kauffmann}, {Heckman}, {White},
  {Charlot}, {Tremonti}, {Peng}, {Seibert}, {Brinkmann}, {Nichol}, {SubbaRao},
  \& {York}}]{kauffmann2003}
{Kauffmann} G. {et~al.}, 2003, \mnras, 341, 54

\bibitem[{{Kroupa}(2001)}]{kroupa2001}
{Kroupa} P., 2001, \mnras, 322, 231

\bibitem[{{Lani} {et~al}\mbox{.}(2013){Lani}, {Almaini}, {Hartley}, {Mortlock},
  {H{\"a}u{\ss}ler}, {Chuter}, {Simpson}, {van der Wel}, {Gr{\"u}tzbauch},
  {Conselice}, {Bradshaw}, {Cooper}, {Faber}, {Grogin}, {Kocevski},
  {Koekemoer}, \& {Lai}}]{lani2013}
{Lani} C. {et~al.}, 2013, \mnras, 435, 207

\bibitem[{{Lilly} {et~al}\mbox{.}(1998){Lilly}, {Schade}, {Ellis}, {Le Fevre},
  {Brinchmann}, {Tresse}, {Abraham}, {Hammer}, {Crampton}, {Colless},
  {Glazebrook}, {Mallen-Ornelas}, \& {Broadhurst}}]{lilly1998}
{Lilly} S. {et~al.}, 1998, \apj, 500, 75

\bibitem[{{Maltby} {et~al}\mbox{.}(2010){Maltby}, {Arag{\'o}n-Salamanca},
  {Gray}, {Barden}, {H{\"a}u{\ss}ler}, {Wolf}, {Peng}, {Jahnke}, {McIntosh},
  {B{\"o}hm}, \& {van Kampen}}]{maltby2010}
{Maltby} D.~T. {et~al.}, 2010, \mnras, 402, 282

\bibitem[{{Maulbetsch} {et~al}\mbox{.}(2007){Maulbetsch}, {Avila-Reese},
  {Col{\'{\i}}n}, {Gottl{\"o}ber}, {Khalatyan}, \&
  {Steinmetz}}]{maulbetsch2007}
{Maulbetsch} C., {Avila-Reese} V., {Col{\'{\i}}n} P., {Gottl{\"o}ber} S.,
  {Khalatyan} A., {Steinmetz} M., 2007, \apj, 654, 53

\bibitem[{{McIntosh} {et~al}\mbox{.}(2005){McIntosh}, {Bell}, {Rix}, {Wolf},
  {Heymans}, {Peng}, {Somerville}, {Barden}, {Beckwith}, {Borch}, {Caldwell},
  {H{\"a}u{\ss}ler}, {Jahnke}, {Jogee}, {Meisenheimer}, {S{\'a}nchez}, \&
  {Wisotzki}}]{mcintosh2005}
{McIntosh} D.~H. {et~al.}, 2005, \apj, 632, 191

\bibitem[{{Muldrew} {et~al}\mbox{.}(2012){Muldrew}, {Croton}, {Skibba},
  {Pearce}, {Ann}, {Baldry}, {Brough}, {Choi}, {Conselice}, {Cowan},
  {Gallazzi}, {Gray}, {Gr{\"u}tzbauch}, {Li}, {Park}, {Pilipenko}, {Podgorzec},
  {Robotham}, {Wilman}, {Yang}, {Zhang}, \& {Zibetti}}]{muldrew2012}
{Muldrew} S.~I. {et~al.}, 2012, \mnras, 419, 2670

\bibitem[{{Nair}, {van den Bergh} \& {Abraham}(2011){Nair}, {van den Bergh}, \&
  {Abraham}}]{nair2011}
{Nair} P., {van den Bergh} S., {Abraham} R.~G., 2011, \apjl, 734, L31

\bibitem[{{Newman} {et~al}\mbox{.}(2012){Newman}, {Ellis}, {Bundy}, \&
  {Treu}}]{newman2012}
{Newman} A.~B., {Ellis} R.~S., {Bundy} K., {Treu} T., 2012, \apj, 746, 162

\bibitem[{{Nipoti}, {Londrillo} \& {Ciotti}(2003){Nipoti}, {Londrillo}, \&
  {Ciotti}}]{nipoti2003}
{Nipoti} C., {Londrillo} P., {Ciotti} L., 2003, \mnras, 342, 501

\bibitem[{{Nipoti} {et~al}\mbox{.}(2009){Nipoti}, {Treu}, {Auger}, \&
  {Bolton}}]{nipoti2009}
{Nipoti} C., {Treu} T., {Auger} M.~W., {Bolton} A.~S., 2009, \apjl, 706, L86

\bibitem[{{Nipoti} {et~al}\mbox{.}(2012){Nipoti}, {Treu}, {Leauthaud}, {Bundy},
  {Newman}, \& {Auger}}]{nipoti2012}
{Nipoti} C., {Treu} T., {Leauthaud} A., {Bundy} K., {Newman} A.~B., {Auger}
  M.~W., 2012, \mnras, 422, 1714

\bibitem[{{Papovich} {et~al}\mbox{.}(2012){Papovich}, {Bassett}, {Lotz}, {van
  der Wel}, {Tran}, {Finkelstein}, {Bell}, {Conselice}, {Dekel}, {Dunlop},
  {Guo}, {Faber}, {Farrah}, {Ferguson}, {Finkelstein}, {H{\"a}ussler},
  {Kocevski}, {Koekemoer}, {Koo}, {McGrath}, {McLure}, {McIntosh}, {Momcheva},
  {Newman}, {Rudnick}, {Weiner}, {Willmer}, \& {Wuyts}}]{papovich2012}
{Papovich} C. {et~al.}, 2012, \apj, 750, 93

\bibitem[{{Poggianti} {et~al}\mbox{.}(2013){Poggianti}, {Calvi}, {Bindoni},
  {D'Onofrio}, {Moretti}, {Valentinuzzi}, {Fasano}, {Fritz}, {De Lucia},
  {Vulcani}, {Bettoni}, {Gullieuszik}, \& {Omizzolo}}]{poggianti2013}
{Poggianti} B.~M. {et~al.}, 2013, \apj, 762, 77

\bibitem[{{Raichoor} {et~al}\mbox{.}(2012){Raichoor}, {Mei}, {Stanford},
  {Holden}, {Nakata}, {Rosati}, {Shankar}, {Tanaka}, {Ford}, {Huertas-Company},
  {Illingworth}, {Kodama}, {Postman}, {Rettura}, {Blakeslee}, {Demarco}, {Jee},
  \& {White}}]{raichoor2012}
{Raichoor} A. {et~al.}, 2012, \apj, 745, 130

\bibitem[{{Ravindranath} {et~al}\mbox{.}(2004){Ravindranath}, {Ferguson},
  {Conselice}, {Giavalisco}, {Dickinson}, {Chatzichristou}, {de Mello}, {Fall},
  {Gardner}, {Grogin}, {Hornschemeier}, {Jogee}, {Koekemoer}, {Kretchmer},
  {Livio}, {Mobasher}, \& {Somerville}}]{ravindranath2004}
{Ravindranath} S. {et~al.}, 2004, \apjl, 604, L9

\bibitem[{{Rettura} {et~al}\mbox{.}(2010){Rettura}, {Rosati}, {Nonino},
  {Fosbury}, {Gobat}, {Menci}, {Strazzullo}, {Mei}, {Demarco}, \&
  {Ford}}]{rettura2010}
{Rettura} A. {et~al.}, 2010, \apj, 709, 512

\bibitem[{{Ricciardelli} {et~al}\mbox{.}(2012){Ricciardelli}, {Vazdekis},
  {Cenarro}, \& {Falc{\'o}n-Barroso}}]{ricciardelli2012}
{Ricciardelli} E., {Vazdekis} A., {Cenarro} A.~J., {Falc{\'o}n-Barroso} J.,
  2012, \mnras, 424, 172

\bibitem[{{Schade} {et~al}\mbox{.}(1996){Schade}, {Lilly}, {Le Fevre},
  {Hammer}, \& {Crampton}}]{schade1996}
{Schade} D., {Lilly} S.~J., {Le Fevre} O., {Hammer} F., {Crampton} D., 1996,
  \apj, 464, 79

\bibitem[{{S\'ersic}(1968)}]{sersic1968}
{S\'ersic} J.~L., 1968, {Atlas de galaxias australes}

\bibitem[{{Shankar} {et~al}\mbox{.}(2010){Shankar}, {Marulli}, {Bernardi},
  {Boylan-Kolchin}, {Dai}, \& {Khochfar}}]{shankar2010b}
{Shankar} F., {Marulli} F., {Bernardi} M., {Boylan-Kolchin} M., {Dai} X.,
  {Khochfar} S., 2010, \mnras, 405, 948

\bibitem[{{Shankar} {et~al}\mbox{.}(2013){Shankar}, {Marulli}, {Bernardi},
  {Mei}, {Meert}, \& {Vikram}}]{shankar2013}
{Shankar} F., {Marulli} F., {Bernardi} M., {Mei} S., {Meert} A., {Vikram} V.,
  2013, \mnras, 428, 109

\bibitem[{{Shankar} {et~al}\mbox{.}(2014){Shankar}, {Mei}, {Huertas-Company},
  {Moreno}, {Fontanot}, {Monaco}, {Bernardi}, {Cattaneo}, {Sheth}, {Licitra},
  {Delaye}, \& {Raichoor}}]{shankar2014}
{Shankar} F. {et~al.}, 2014, \mnras, 439, 3189

\bibitem[{{Shen} {et~al}\mbox{.}(2003){Shen}, {Mo}, {White}, {Blanton},
  {Kauffmann}, {Voges}, {Brinkmann}, \& {Csabai}}]{shen2003}
{Shen} S., {Mo} H.~J., {White} S.~D.~M., {Blanton} M.~R., {Kauffmann} G.,
  {Voges} W., {Brinkmann} J., {Csabai} I., 2003, \mnras, 343, 978

\bibitem[{{Sheth} \& {Tormen}(2004)}]{sheth2004}
{Sheth} R.~K., {Tormen} G., 2004, \mnras, 350, 1385

\bibitem[{{Simard} {et~al}\mbox{.}(1999){Simard}, {Koo}, {Faber}, {Sarajedini},
  {Vogt}, {Phillips}, {Gebhardt}, {Illingworth}, \& {Wu}}]{simard1999}
{Simard} L. {et~al.}, 1999, \apj, 519, 563

\bibitem[{{Strauss} {et~al}\mbox{.}(2002){Strauss}, {Weinberg}, {Lupton},
  {Narayanan}, {Annis}, {Bernardi}, {Blanton}, {Burles}, {Connolly},
  {Dalcanton}, {Doi}, {Eisenstein}, {Frieman}, {Fukugita}, {Gunn},
  {Ivezi{\'c}}, {Kent}, {Kim}, {Knapp}, {Kron}, {Munn}, {Newberg}, {Nichol},
  {Okamura}, {Quinn}, {Richmond}, {Schlegel}, {Shimasaku}, {SubbaRao},
  {Szalay}, {Vanden Berk}, {Vogeley}, {Yanny}, {Yasuda}, {York}, \&
  {Zehavi}}]{strauss2002}
{Strauss} M.~A. {et~al.}, 2002, \aj, 124, 1810

\bibitem[{{Strazzullo} {et~al}\mbox{.}(2013){Strazzullo}, {Gobat}, {Daddi},
  {Onodera}, {Carollo}, {Dickinson}, {Renzini}, {Arimoto}, {Cimatti},
  {Finoguenov}, \& {Chary}}]{strazzullo2013}
{Strazzullo} V. {et~al.}, 2013, \apj, 772, 118

\bibitem[{{Stringer} {et~al}\mbox{.}(2013){Stringer}, {Shankar}, {Novak},
  {Huertas-Company}, {Combes}, \& {Moster}}]{stringer2013}
{Stringer} M.~J., {Shankar} F., {Novak} G.~S., {Huertas-Company} M., {Combes}
  F., {Moster} B.~P., 2013, ArXiv, 1310.3823

\bibitem[{{Szabo} {et~al}\mbox{.}(2011){Szabo}, {Pierpaoli}, {Dong}, {Pipino},
  \& {Gunn}}]{szabo2011}
{Szabo} T., {Pierpaoli} E., {Dong} F., {Pipino} A., {Gunn} J., 2011, \apj, 736,
  21

\bibitem[{{Szomoru}, {Franx} \& {van Dokkum}(2012){Szomoru}, {Franx}, \& {van
  Dokkum}}]{szomoru2012}
{Szomoru} D., {Franx} M., {van Dokkum} P.~G., 2012, \apj, 749, 121

\bibitem[{{Takey}, {Schwope} \& {Lamer}(2011){Takey}, {Schwope}, \&
  {Lamer}}]{takey2011}
{Takey} A., {Schwope} A., {Lamer} G., 2011, \aap, 534, A120

\bibitem[{{Taylor} {et~al}\mbox{.}(2010){Taylor}, {Franx}, {Glazebrook},
  {Brinchmann}, {van der Wel}, \& {van Dokkum}}]{taylor2010}
{Taylor} E.~N., {Franx} M., {Glazebrook} K., {Brinchmann} J., {van der Wel} A.,
  {van Dokkum} P.~G., 2010, \apj, 720, 723

\bibitem[{{Tempel}, {Tago} \& {Liivam{\"a}gi}(2012){Tempel}, {Tago}, \&
  {Liivam{\"a}gi}}]{tempel2012}
{Tempel} E., {Tago} E., {Liivam{\"a}gi} L.~J., 2012, \aap, 540, A106

\bibitem[{{Trujillo} \& {Aguerri}(2004)}]{trujilloaguerri2004}
{Trujillo} I., {Aguerri} J.~A.~L., 2004, \mnras, 355, 82

\bibitem[{{Trujillo} {et~al}\mbox{.}(2009){Trujillo}, {Cenarro}, {de
  Lorenzo-C{\'a}ceres}, {Vazdekis}, {de la Rosa}, \& {Cava}}]{trujillo2009}
{Trujillo} I., {Cenarro} A.~J., {de Lorenzo-C{\'a}ceres} A., {Vazdekis} A., {de
  la Rosa} I.~G., {Cava} A., 2009, \apjl, 692, L118

\bibitem[{{Trujillo} {et~al}\mbox{.}(2007){Trujillo}, {Conselice}, {Bundy},
  {Cooper}, {Eisenhardt}, \& {Ellis}}]{trujillo2007}
{Trujillo} I., {Conselice} C.~J., {Bundy} K., {Cooper} M.~C., {Eisenhardt} P.,
  {Ellis} R.~S., 2007, \mnras, 382, 109

\bibitem[{{Trujillo} {et~al}\mbox{.}(2006){Trujillo}, {F{\"o}rster Schreiber},
  {Rudnick}, {Barden}, {Franx}, {Rix}, {Caldwell}, {McIntosh}, {Toft},
  {H{\"a}ussler}, {Zirm}, {van Dokkum}, {Labb{\'e}}, {Moorwood},
  {R{\"o}ttgering}, {van der Wel}, {van der Werf}, \& {van
  Starkenburg}}]{trujillo2006}
{Trujillo} I. {et~al.}, 2006, \apj, 650, 18

\bibitem[{{Trujillo} {et~al}\mbox{.}(2004){Trujillo}, {Rudnick}, {Rix},
  {Labb{\'e}}, {Franx}, {Daddi}, {van Dokkum}, {F{\"o}rster Schreiber},
  {Kuijken}, {Moorwood}, {R{\"o}ttgering}, {van der Wel}, {van der Werf}, \&
  {van Starkenburg}}]{trujillo2004}
{Trujillo} I. {et~al.}, 2004, \apj, 604, 521

\bibitem[{{Valentinuzzi} {et~al}\mbox{.}(2010){Valentinuzzi}, {Fritz},
  {Poggianti}, {Cava}, {Bettoni}, {Fasano}, {D'Onofrio}, {Couch}, {Dressler},
  {Moles}, {Moretti}, {Omizzolo}, {Kj{\ae}rgaard}, {Vanzella}, \&
  {Varela}}]{valentinuzzi2010}
{Valentinuzzi} T. {et~al.}, 2010, \apj, 712, 226

\bibitem[{{van der Wel} {et~al}\mbox{.}(2005){van der Wel}, {Franx}, {van
  Dokkum}, {Rix}, {Illingworth}, \& {Rosati}}]{vanderwel2005}
{van der Wel} A., {Franx} M., {van Dokkum} P.~G., {Rix} H.-W., {Illingworth}
  G.~D., {Rosati} P., 2005, \apj, 631, 145

\bibitem[{{van der Wel} {et~al}\mbox{.}(2014){van der Wel}, {Franx}, {van
  Dokkum}, {Skelton}, {Momcheva}, {Whitaker}, {Brammer}, {Bell}, {Rix},
  {Wuyts}, {Ferguson}, {Holden}, {Barro}, {Koekemoer}, {Chang}, {McGrath},
  {Haussler}, {Dekel}, {Behroozi}, {Fumagalli}, {Leja}, {Lundgren}, {Maseda},
  {Nelson}, {Wake}, {Patel}, {Labbe}, {Faber}, {Grogin}, \&
  {Kocevski}}]{vanderwel2014}
{van der Wel} A. {et~al.}, 2014, ArXiv, 1404.2844

\bibitem[{{Varela} {et~al}\mbox{.}(2012){Varela}, {Betancort-Rijo}, {Trujillo},
  \& {Ricciardelli}}]{varela2012}
{Varela} J., {Betancort-Rijo} J., {Trujillo} I., {Ricciardelli} E., 2012, \apj,
  744, 82

\bibitem[{{Varela} {et~al}\mbox{.}(2009){Varela}, {D'Onofrio}, {Marmo},
  {Fasano}, {Bettoni}, {Cava}, {Couch}, {Dressler}, {Kj{\ae}rgaard}, {Moles},
  {Pignatelli}, {Poggianti}, \& {Valentinuzzi}}]{varela2009}
{Varela} J. {et~al.}, 2009, \aap, 497, 667

\bibitem[{{Vazdekis} {et~al}\mbox{.}(2012){Vazdekis}, {Ricciardelli},
  {Cenarro}, {Rivero-Gonz{\'a}lez}, {D{\'{\i}}az-Garc{\'{\i}}a}, \&
  {Falc{\'o}n-Barroso}}]{vazdekis2012}
{Vazdekis} A., {Ricciardelli} E., {Cenarro} A.~J., {Rivero-Gonz{\'a}lez} J.~G.,
  {D{\'{\i}}az-Garc{\'{\i}}a} L.~A., {Falc{\'o}n-Barroso} J., 2012, \mnras,
  424, 157

\bibitem[{{Verdes-Montenegro} {et~al}\mbox{.}(2005){Verdes-Montenegro},
  {Sulentic}, {Lisenfeld}, {Leon}, {Espada}, {Garcia}, {Sabater}, \&
  {Verley}}]{2005verdes-montenegro}
{Verdes-Montenegro} L., {Sulentic} J., {Lisenfeld} U., {Leon} S., {Espada} D.,
  {Garcia} E., {Sabater} J., {Verley} S., 2005, \aap, 436, 443

\bibitem[{{Wen}, {Han} \& {Liu}(2012){Wen}, {Han}, \& {Liu}}]{wen2012}
{Wen} Z.~L., {Han} J.~L., {Liu} F.~S., 2012, \apjs, 199, 34

\end{thebibliography}

}

\appendix

\section[]{Numerical values for means and dispersions of the stellar mass-size relations of this work}
\label{sec:appendixA}

\begin{table*}
    \begin{tabular}{c|cccccc}
      \hline
	Mass    &All & Underdense & Overdense  &All & 10\% Underdense & 10\% Overdense\\  
($\times10^{10}M_{\sun}$)     & $\overline{R}_e$ (kpc) & $\overline{R}_e$ (kpc) & $\overline{R}_e$ (kpc) & $\sigma_{\ln(R_e)}$ & $\sigma_{\ln(R_e)}$ & $\sigma_{\ln(R_e)}$\\
      \hline
 $0.2$  &  $1.71\pm0.02$   & $1.66\pm0.05$   & $1.71\pm0.05$ &  $0.490\pm0.005$   & $0.517\pm0.01$   & $0.421\pm0.02$  \\
 $0.4$  &  $2.09\pm0.01$   & $2.09\pm0.05$   & $1.90\pm0.04$ &  $0.457\pm0.005$   & $0.457\pm0.01$   & $0.394\pm0.01$  \\
 $0.7$  &  $2.33\pm0.02$   & $2.33\pm0.05$   & $2.09\pm0.04$ &  $0.454\pm0.004$   & $0.475\pm0.01$   & $0.403\pm0.01$  \\
 $1.0$  &  $2.52\pm0.01$   & $2.52\pm0.04$   & $2.28\pm0.06$ &  $0.436\pm0.004$   & $0.439\pm0.01$   & $0.421\pm0.01$  \\
 $1.3$  &  $2.76\pm0.03$   & $2.80\pm0.05$   & $2.52\pm0.06$ &  $0.427\pm0.005$   & $0.442\pm0.01$   & $0.385\pm0.01$  \\
 $1.6$  &  $2.85\pm0.02$   & $2.90\pm0.05$   & $2.66\pm0.06$ &  $0.406\pm0.004$   & $0.397\pm0.010$  & $0.400\pm0.01$  \\
 $1.9$  &  $3.04\pm0.02$   & $2.99\pm0.05$   & $2.80\pm0.06$ &  $0.394\pm0.004$   & $0.391\pm0.01$   & $0.391\pm0.01$  \\
 $2.2$  &  $3.18\pm0.03$   & $3.18\pm0.05$   & $2.95\pm0.06$ &  $0.388\pm0.004$   & $0.391\pm0.01$   & $0.373\pm0.01$  \\
 $2.5$  &  $3.33\pm0.03$   & $3.28\pm0.06$   & $3.18\pm0.06$ &  $0.379\pm0.004$   & $0.388\pm0.010$  & $0.370\pm0.01$  \\
 $2.9$  &  $3.52\pm0.03$   & $3.52\pm0.05$   & $3.37\pm0.06$ &  $0.358\pm0.003$   & $0.352\pm0.010$  & $0.355\pm0.01$  \\
 $3.3$  &  $3.66\pm0.04$   & $3.61\pm0.05$   & $3.52\pm0.07$ &  $0.355\pm0.004$   & $0.358\pm0.009$  & $0.361\pm0.01$  \\
 $3.8$  &  $3.85\pm0.02$   & $3.85\pm0.06$   & $3.71\pm0.05$ &  $0.346\pm0.004$   & $0.346\pm0.009$  & $0.361\pm0.010$ \\
 $4.3$  &  $4.04\pm0.03$   & $4.04\pm0.05$   & $3.90\pm0.06$ &  $0.340\pm0.004$   & $0.322\pm0.008$  & $0.355\pm0.01$  \\
 $5.3$  &  $4.28\pm0.02$   & $4.28\pm0.07$   & $4.09\pm0.06$ &  $0.334\pm0.003$   & $0.325\pm0.009$  & $0.334\pm0.009$ \\
 $7.5$  &  $4.85\pm0.02$   & $4.85\pm0.06$   & $4.89\pm0.07$ &  $0.310\pm0.003$   & $0.313\pm0.009$  & $0.337\pm0.01$  \\

      \hline
        \end{tabular}
   \caption{The mean effective radii and the scatter of the effective radii for late-type galaxies ($n<2.5$) when using the mass density around the galaxies as an environment indicator.}
     \label{table:lateneigh}
\end{table*}

\begin{table*}
    \begin{tabular}{c|cccccc}
      \hline
	Mass    &All & Underdense & Overdense  &All & 10\% Underdense & 10\% Overdense\\  
($\times10^{10}M_{\sun}$)     & $\overline{R}_e$ (kpc) & $\overline{R}_e$ (kpc) & $\overline{R}_e$ (kpc) & $\sigma_{\ln(R_e)}$ & $\sigma_{\ln(R_e)}$ & $\sigma_{\ln(R_e)}$\\
      \hline
 $0.9$   &  $1.38\pm0.01$   & $1.33\pm0.04$   & $1.38\pm0.03$ &  $0.487\pm0.004$   & $0.478\pm0.01$    & $0.466\pm0.010$ \\
 $1.6$   &  $1.76\pm0.02$   & $1.76\pm0.03$   & $1.76\pm0.03$ &  $0.463\pm0.004$   & $0.481\pm0.01$    & $0.436\pm0.009$ \\
 $2.1$   &  $2.00\pm0.03$   & $2.00\pm0.04$   & $1.95\pm0.04$ &  $0.457\pm0.003$   & $0.457\pm0.01$    & $0.442\pm0.009$ \\
 $2.6$   &  $2.23\pm0.02$   & $2.19\pm0.03$   & $2.14\pm0.04$ &  $0.448\pm0.004$   & $0.448\pm0.01$    & $0.427\pm0.010$ \\
 $3.0$   &  $2.42\pm0.01$   & $2.47\pm0.04$   & $2.38\pm0.05$ &  $0.439\pm0.003$   & $0.454\pm0.010$   & $0.421\pm0.008$ \\
 $3.4$   &  $2.66\pm0.03$   & $2.61\pm0.04$   & $2.57\pm0.04$ &  $0.436\pm0.004$   & $0.427\pm0.009$   & $0.421\pm0.009$ \\
 $3.8$   &  $2.80\pm0.03$   & $2.76\pm0.06$   & $2.71\pm0.04$ &  $0.421\pm0.004$   & $0.454\pm0.01$    & $0.397\pm0.008$ \\
 $4.2$   &  $2.99\pm0.03$   & $2.99\pm0.04$   & $2.95\pm0.05$ &  $0.409\pm0.003$   & $0.403\pm0.009$   & $0.397\pm0.008$ \\
 $4.7$   &  $3.23\pm0.03$   & $3.23\pm0.05$   & $3.18\pm0.05$ &  $0.406\pm0.003$   & $0.418\pm0.009$   & $0.388\pm0.008$ \\
 $5.2$   &  $3.42\pm0.03$   & $3.37\pm0.06$   & $3.37\pm0.05$ &  $0.388\pm0.003$   & $0.397\pm0.010$   & $0.391\pm0.008$ \\
 $5.9$   &  $3.71\pm0.01$   & $3.71\pm0.06$   & $3.61\pm0.04$ &  $0.379\pm0.003$   & $0.379\pm0.008$   & $0.379\pm0.008$ \\
 $6.8$   &  $4.04\pm0.02$   & $4.04\pm0.06$   & $4.04\pm0.05$ &  $0.364\pm0.003$   & $0.382\pm0.009$   & $0.367\pm0.007$ \\
 $7.9$   &  $4.42\pm0.02$   & $4.32\pm0.07$   & $4.42\pm0.05$ &  $0.352\pm0.003$   & $0.355\pm0.008$   & $0.352\pm0.007$ \\
 $9.8$   &  $5.04\pm0.02$   & $4.99\pm0.07$   & $5.08\pm0.06$ &  $0.328\pm0.003$   & $0.355\pm0.009$   & $0.325\pm0.007$ \\
 $14.5$  &  $6.51\pm0.04$   & $6.41\pm0.09$   & $6.84\pm0.05$ &  $0.316\pm0.002$   & $0.322\pm0.009$   & $0.322\pm0.005$ \\
      \hline
        \end{tabular}
   \caption{Same as Table \ref{table:lateneigh} but for early-type galaxies ($n>2.5$).}
     \label{table:earlyneigh}
\end{table*}

\begin{table*}
    \begin{tabular}{c|cccccc}
      \hline
	Mass    &All & Field & Clusters &All & Field & Clusters\\  
($\times10^{10}M_{\sun}$)     & $\overline{R}_e$ (kpc) & $\overline{R}_e$ (kpc) & $\overline{R}_e$ (kpc) & $\sigma_{\ln(R_e)}$ & $\sigma_{\ln(R_e)}$ & $\sigma_{\ln(R_e)}$\\
      \hline
 $0.2$  &  $1.71\pm0.02$   & $1.71\pm0.02$   & $1.66\pm0.06$ &  $0.490\pm0.005$   & $0.493\pm0.005$   & $0.433\pm0.02$ \\
 $0.4$  &  $2.09\pm0.01$   & $2.09\pm0.03$   & $1.81\pm0.05$ &  $0.457\pm0.005$   & $0.460\pm0.005$   & $0.385\pm0.01$ \\
 $0.7$  &  $2.33\pm0.02$   & $2.33\pm0.02$   & $2.09\pm0.05$ &  $0.454\pm0.004$   & $0.454\pm0.005$   & $0.415\pm0.01$ \\
 $1.0$  &  $2.52\pm0.01$   & $2.52\pm0.02$   & $2.28\pm0.06$ &  $0.436\pm0.004$   & $0.439\pm0.005$   & $0.385\pm0.02$ \\
 $1.3$  &  $2.76\pm0.03$   & $2.76\pm0.02$   & $2.52\pm0.06$ &  $0.427\pm0.005$   & $0.427\pm0.005$   & $0.418\pm0.02$ \\
 $1.6$  &  $2.85\pm0.02$   & $2.85\pm0.04$   & $2.66\pm0.06$ &  $0.406\pm0.004$   & $0.406\pm0.004$   & $0.406\pm0.02$ \\
 $1.9$  &  $3.04\pm0.02$   & $3.04\pm0.02$   & $2.80\pm0.07$ &  $0.394\pm0.004$   & $0.394\pm0.004$   & $0.379\pm0.01$ \\
 $2.2$  &  $3.18\pm0.03$   & $3.18\pm0.02$   & $2.95\pm0.08$ &  $0.388\pm0.004$   & $0.385\pm0.004$   & $0.400\pm0.02$ \\
 $2.5$  &  $3.33\pm0.03$   & $3.37\pm0.03$   & $3.09\pm0.07$ &  $0.379\pm0.004$   & $0.379\pm0.004$   & $0.379\pm0.02$ \\
 $2.9$  &  $3.52\pm0.03$   & $3.52\pm0.02$   & $3.33\pm0.08$ &  $0.358\pm0.003$   & $0.358\pm0.003$   & $0.358\pm0.02$ \\
 $3.3$  &  $3.66\pm0.04$   & $3.71\pm0.03$   & $3.37\pm0.08$ &  $0.355\pm0.004$   & $0.355\pm0.004$   & $0.355\pm0.02$ \\
 $3.8$  &  $3.85\pm0.02$   & $3.85\pm0.03$   & $3.52\pm0.1$  &  $0.346\pm0.004$   & $0.343\pm0.003$   & $0.424\pm0.02$ \\
 $4.3$  &  $4.04\pm0.03$   & $4.04\pm0.02$   & $3.80\pm0.09$ &  $0.340\pm0.004$   & $0.340\pm0.004$   & $0.343\pm0.02$ \\
 $5.3$  &  $4.28\pm0.02$   & $4.28\pm0.02$   & $4.13\pm0.10$ &  $0.334\pm0.003$   & $0.334\pm0.003$   & $0.334\pm0.01$ \\
 $7.5$  &  $4.85\pm0.02$   & $4.85\pm0.02$   & $4.80\pm0.1$  &  $0.310\pm0.003$   & $0.307\pm0.004$   & $0.355\pm0.02$ \\
      \hline
        \end{tabular}
   \caption{The mean effective radii and the scatter of the effective radii for late-type galaxies ($n<2.5$) depending whether they inhabit a galaxy cluster or not.}
     \label{table:lateclust}
\end{table*}

\begin{table*}
    \begin{tabular}{c|cccccc}
      \hline
	Mass    &All & Field & Clusters &All & Field & Clusters\\  
	($\times10^{10}M_{\sun}$)     & $\overline{R}_e$ (kpc) & $\overline{R}_e$ (kpc) & $\overline{R}_e$ (kpc) & $\sigma_{\ln(R_e)}$ & $\sigma_{\ln(R_e)}$ & $\sigma_{\ln(R_e)}$\\
      \hline
 $0.9$  &  $1.38\pm0.01$   & $1.38\pm0.01$   & $1.43\pm0.02$ &  $0.487\pm0.004$   & $0.493\pm0.004$   & $0.424\pm0.01$ \\
 $1.6$  &  $1.76\pm0.02$   & $1.76\pm0.03$   & $1.71\pm0.04$ &  $0.463\pm0.004$   & $0.463\pm0.004$   & $0.442\pm0.01$ \\
 $2.1$  &  $2.00\pm0.03$   & $2.04\pm0.03$   & $1.90\pm0.04$ &  $0.457\pm0.003$   & $0.460\pm0.003$   & $0.415\pm0.01$ \\
 $2.6$  &  $2.23\pm0.02$   & $2.23\pm0.02$   & $2.19\pm0.05$ &  $0.448\pm0.004$   & $0.451\pm0.004$   & $0.433\pm0.01$ \\
 $3.0$  &  $2.42\pm0.01$   & $2.42\pm0.01$   & $2.33\pm0.05$ &  $0.439\pm0.003$   & $0.439\pm0.003$   & $0.436\pm0.01$ \\
 $3.4$  &  $2.66\pm0.03$   & $2.66\pm0.03$   & $2.57\pm0.05$ &  $0.436\pm0.004$   & $0.436\pm0.003$   & $0.427\pm0.01$ \\
 $3.8$  &  $2.80\pm0.03$   & $2.80\pm0.02$   & $2.66\pm0.04$ &  $0.421\pm0.004$   & $0.421\pm0.003$   & $0.385\pm0.01$ \\
 $4.2$  &  $2.99\pm0.03$   & $2.99\pm0.03$   & $2.85\pm0.06$ &  $0.409\pm0.003$   & $0.409\pm0.003$   & $0.382\pm0.01$ \\
 $4.7$  &  $3.23\pm0.03$   & $3.23\pm0.03$   & $3.09\pm0.06$ &  $0.406\pm0.003$   & $0.406\pm0.003$   & $0.403\pm0.01$ \\
 $5.2$  &  $3.42\pm0.03$   & $3.47\pm0.03$   & $3.28\pm0.06$ &  $0.388\pm0.003$   & $0.388\pm0.004$   & $0.367\pm0.01$ \\
 $5.9$  &  $3.71\pm0.01$   & $3.71\pm0.02$   & $3.61\pm0.05$ &  $0.379\pm0.003$   & $0.379\pm0.003$   & $0.370\pm0.01$ \\
 $6.8$  &  $4.04\pm0.02$   & $4.04\pm0.01$   & $3.85\pm0.08$ &  $0.364\pm0.003$   & $0.361\pm0.004$   & $0.388\pm0.01$ \\
 $7.9$  &  $4.42\pm0.02$   & $4.42\pm0.02$   & $4.37\pm0.07$ &  $0.352\pm0.003$   & $0.352\pm0.003$   & $0.352\pm0.01$ \\
 $9.8$  &  $5.04\pm0.02$   & $5.04\pm0.03$   & $4.94\pm0.08$ &  $0.328\pm0.003$   & $0.328\pm0.004$   & $0.322\pm0.010$ \\
 $14.5$ &  $6.51\pm0.04$   & $6.51\pm0.03$   & $6.89\pm0.09$ &  $0.316\pm0.002$   & $0.313\pm0.004$   & $0.331\pm0.008$ \\
      \hline
        \end{tabular}
   \caption{Same as Table \ref{table:lateclust} but for early-type galaxies.}
     \label{table:earlyclust}
\end{table*}

\section[]{Influence of observational errors in the observed scatter}
\label{sec:appendixB}

To have a hint of the effect of the observational errors on the observed scatter of the stellar mass-size relations (see Section \ref{sec:discussion}), we conduct a simple experiment using the mean sizes at a fixed stellar mass obtained in Section \ref{sec:size-scatter} for the whole sample. We consider that the errors on the measurements of the stellar mass ($\frac{\delta \log M_{*}}{\log M_{*}}$) and sizes ($\frac{\delta R_e}{R_e}$) of the galaxies are gaussianly distributed. The different errors used (optimistic, realistic, pessimistic scenarios) are:

\begin{equation} \label{eq:errors_mass} 
\frac{\delta \log M_{*}}{\log M_{*}}=0.1,0.2,0.25 
\end{equation}

\begin{equation} \label{eq:errors_size} 
\frac{\delta \overline{R}_e}{\overline{R}_e}=0.10,0.15,0.20
\end{equation}

We generate 5000 random mock galaxies following the gaussian distributions given by the errors with the mean in stellar mass and size corresponding to the measured values. On doing this, we simulate the scatter around the stellar mass-size relation produced by the observational errors for both late and early-type galaxies in the clusters, in the field and in the whole sample. Finally, we apply the maximum likelihood method detailed in Section \ref{sec:par-estimation} to the mock samples. This allows us to obtain the scatter values corresponding to a sample only dominated by errors. It is worth noting that, since the mean size and stellar mass values are an input to create the mock sample, we recover the same values as an output of the likelihood analysis.

\begin{figure*}
  \includegraphics[width=168mm]{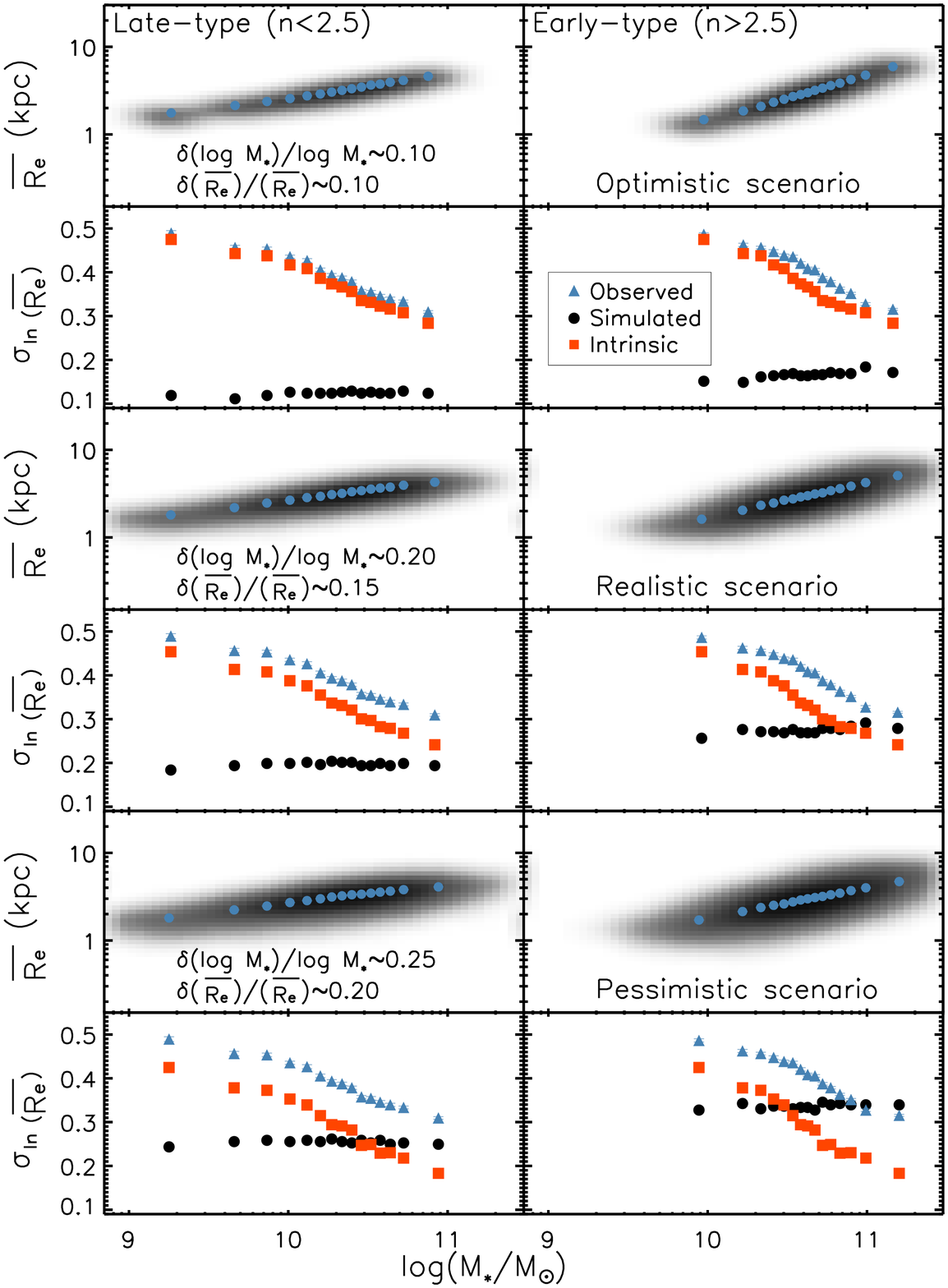}
 \caption{Simulated stellar mass-size distribution and scatter using three different scenarios for the error in stellar mass and the error in the effective radius. Upper panels for each case show the distribution (shaded surface) and the stellar mass-size relation used to generate it (blue filled dots). Lower panels compare the observed scatter in the relation (blue triangles) and the scatter when only the errors have been taken into account (black filled dots). The resulting 'intrinsic' scatter is shown as filled dark-orange squares.}
  \label{fig:errors_mock}
\end{figure*}

Figure \ref{fig:errors_mock} shows the resulting distribution and the measured scatter for different error values, both in stellar mass and in size, as well as the observed and intrinsic scatter (i.e. the corrected value accounting for the error simulation) for different values of the observational errors. It is interesting to note how at higher stellar masses the observed scatter becomes dominated by errors in the second and third cases (middle and lower panels). In fact, Figure \ref{fig:errors_mock} shows a larger scatter in the pessimistic scenario for high stellar masses than that observed in this work, probably because the assumed errors in this simulation is larger than the errors in our sample.

\bsp

\label{lastpage}

\end{document}